\documentclass[12pt,preprint]{aastex}

\shorttitle{Evolution of Global Relativistic Jets}
\shortauthors{Nishikawa et al.}

\begin{document}


\title{Evolution of Global Relativistic Jets: \\
    Collimations and Expansion with kKHI and the Weibel Instability}



\author{K.-I. Nishikawa\altaffilmark{1}, J. T. Frederiksen\altaffilmark{2},  \AA. Nordlund\altaffilmark{2},  
Y. Mizuno\altaffilmark{3}, P.E. Hardee\altaffilmark{4}, \\ J. Niemiec\altaffilmark{5}, J. L. G\'{o}mez\altaffilmark{6}, 
A. Pe'er\altaffilmark{7},
I. Du\c{t}an\altaffilmark{8}, 
A. Meli\altaffilmark{9}, H. Sol\altaffilmark{10},  M. Pohl\altaffilmark{11,12}, \\
D. H. Hartmann\altaffilmark{13}}
  
\altaffiltext{1}{Department of Physics,
University of Alabama in Huntsville, ZP12,
Huntsville, AL 35899, USA; ken-ichi.nishikawa@nasa.gov}
 
\altaffiltext{2}{Niels Bohr Institute, University of Copenhagen, Blegdamsvej 17, DK-2100 Copenhagen, Denmark}
 
\altaffiltext{3}{Institute for Theoretical Physics, Goethe University, D-60438, Frankfurt am Main, Germany}

\altaffiltext{4}{Department of Physics and Astronomy, The University
of Alabama, Tuscaloosa, AL 35487, USA} 

\altaffiltext{5}{Institute of Nuclear Physics PAN, ul. Radzikowskiego
152, 31-342 Krak\'{o}w, Poland}

\altaffiltext{6} {Instituto de Astrof\'{i}sica de Andaluc\'{i}a, CSIC, Apartado 3004, 18080 Granada, Spain}
\altaffiltext{7}{Physics Department, University College Cork, Cork, Ireland}
\altaffiltext{8}{Institute of Space Science, Atomistilor 409, Bucharest-Magurele RO-077125, Romania}
  
\altaffiltext{9}{Department  of Physics and Astronomy,
University of Gent, Proeftuinstraat 86 B-9000, 
Gent, Belgium}
 
\altaffiltext{10}{LUTH, Observatore de Paris-Meudon, 5 place Jules
Jansen, 92195 Meudon Cedex, France}
      
\altaffiltext{11}{Institut fur Physik und Astronomie, Universit\"{a}t Potsdam, 14476 Potsdam-Golm, Germany}
\altaffiltext{12}{DESY, Platanenallee 6, 15738 Zeuthen, Germany}
\altaffiltext{13}{Department of Physics and Astronomy, Clemson University, Clemson, SC 29634, USA}


\begin{abstract}

In the study of relativistic jets one of the key open questions is their interaction with the environment. Here, we study the initial evolution of both electron-proton ($e^{-}-p^{+}$) and electron-positron ($e^{\pm}$) relativistic jets, focusing on their lateral interaction with ambient plasma. We follow the evolution of toroidal magnetic fields generated by both the kinetic Kelvin-Helmholtz (kKH) and Mushroom instabilities (MI). For an $e^{-}-p^{+}$  jet, the induced magnetic field collimates the jet and electrons are perpendicularly accelerated. As the instabilities saturate and subsequently weaken, the magnetic polarity switches from clockwise to counter-clockwise in the middle of the jet. For an $e^{\pm}$ jet, we find strong mixing of electrons and positrons with the ambient plasma, resulting in the creation of a bow shock. The merging of current filaments generates density inhomogeneities which initiate a forward shock. 
Strong jet-ambient plasma mixing prevents a full development of the jet (on the scale studied), revealing evidence for both jet collimation and particle acceleration in the forming bow shock. Differences in the magnetic field structure generated by $e^{-}-p^{+}$ and $e^{\pm}$ jets may  contribute to the polarization properties of the observed emission in AGN jets and gamma ray bursts.

\end{abstract}


\section{Introduction}
\vspace{-0.2cm}

Relativistic jets are collimated plasma outflows associated with  active galactic nuclei (AGNs), gamma-ray bursts (GRBs), 
and pulsars.  Among these astrophysical systems,  blazars and GRB jets produce the most luminous phenomena in the universe (e.g., Pe'er 2014). Despite extensive observational and theoretical investigations including
simulation studies, our understanding of their formation, evolution in ambient plasmas, and consequently, their observable properties such as time-dependent flux and polarity, remain quite limited. One of the key open questions in the study of relativistic jets is how they interact with the immediate plasma environment.

The above mentioned outflows are commonly thought to be dynamically hot (relativistic) magnetized plasma flows launched, accelerated, and collimated in regions where Poynting flux dominates over particle (matter) flux (e.g., Blandford \& Znajek 1977; McKinney et al. 2014). This scenario involves a helical large-scale magnetic field structure in some AGN jets, which provides a unique signature in the form of observed asymmetries across the jet width, particularly in the polarization (e.g., Laing 1981; Aloy et al. 2000;  Clausen-Brown, Lyutikov, \& Karb 2011). 

Outflows interact with the interstellar medium and can thereby generate relativistic shocks, as demonstrated in 
particle-in-cell (PIC) simulations (e.g., Nishikawa et al. 2009a) which show that particles are accelerated and radiate predominantly from the magnetic field enhanced reverse shock region. The shocks are collisionless and  result from the growth of  kinetic beam-plasma instabilities: either electrostatic (e.g., two-stream or Buneman modes), quasi electrostatic (e.g., Bret et al. 2010), or electromagnetic (e.g., filamentation). Extensive PIC simulations have been performed to investigate the microscopic processes of jet-driven collisionless relativistic shocks without implementing jet boundaries. The simulations have shown that in shocks in un- or weakly magnetized plasmas beam-plasma instabilities produce current filaments and associated magnetic fields that lead to particle acceleration and emission (e.g., Weibel 1959; Medvedev \& Loeb 1999; Frederiksen et al. 2004; Nishikawa et al. 2003, 2005, 2006, 2008, 2009a; Hededal et al. 2004; Hededal \& Nishikawa 2005; Silva et al. 2003; Jaroschek et al. 2005; Chang et al. 2008; Dieckmann et al. 2008; Spitkovsky 2008a, 2008b; Martins et al. 2009; Sironi \& Spitkovsky 2009a; Haugb\o lle 2011; Sironi et al. 2013; Choi et al. 2014; Ardaneh et al. 2015). 

At the leading edge of the jet, Ardaneh et al. (2015) reported  efficient particle acceleration in a double shock system like that found earlier by Nishikawa et al. (2009a) and subsequently by Choi et al. (2014). The double shock system is comprised of three regions: an acceleration region behind the reverse shock (RS) with strong transverse electromagnetic fields accompanying an ambipolar electrostatic field in the $e^{-} - p^{+}$ case, a heated region with relatively weak electromagnetic fields ahead of the forward shock (FS), and a hot shocked region with large electromagnetic field structures that are of the order 1-30 ion skin depths in size. The simulation captured a fully developed RS while the FS  was still evolving. In order to also investigate velocity shear at flow boundaries, numerous PIC simulations have been performed, demonstrating the growth of the electron-scale Kelvin-Helmholtz instability (ESKHI), also referred to as the kinetic Kelvin-Helmholtz instability (kKHI), e.g., Alves et al. 2012, 2014; Grismayer et al. 2013a, 2013b; Liang et al. 2013a, 2013b; Nishikawa et al. 2013, 2014a,b,c). 

Recently, Alves et al. (2015) reported electron-scale surface waves  in the transverse plane of a counter-streaming sheared flow in an initially unmagnetized collisionless plasma. They labeled  this  the Mushroom instability (MI) due to the Mushroom-like structures that emerge in the electron density. While the ESKHI (kKHI) has higher growth rates than the MI for subrelativistic cases, the MI growth rate scales with $\gamma_{\rm jt}^{-1/2}$ and declines slower than the ESKHI, which scales with
$\gamma_{\rm jt}^{-3/2}$ (Alves et al. 2015). Here $\gamma_{\rm jt}$ is the Lorentz factor
of  jet particles. Nishikawa et al. (2014b) also found that the growth of  transverse structure seen in their 3D simulations likely grows on timescales $t \propto  \gamma_{\rm jt}^{1/2}$. Fluctuation wavelengths along the flow direction seen in lower Lorentz factor simulations are on the order of the predicted fastest growing wavelengths for both electron-proton and electron-positron plasmas. This suggests that the dispersion relation valid for electron-proton plasmas can be applied approximately even for equal mass negatively and positively charged particles. On the other hand, the absolute rate of growth and non-linear structure are very different for electron-proton 
and electron-positron plasmas.

Nishikawa et al. (2014a,b,c) performed three-dimensional (3D) PIC simulations to investigate the  kKHI and MI
using a relativistic jet core with $\gamma_{\rm jt}= 1.5,~5,~15$ and surrounding stationary sheath plasma 
for two different electron-proton  ($e^{-}-p^{+}$)  and electron-positron ($e^{\pm}$) compositions. 
This more physically realistic jet and stationary sheath setup with the slab model allows for spatial propagation and 
provides a proper observer frame view of the shear layer structures.   
Their key findings can be summarized as follows: The $e^{-}-p^{+}$  cases generate a DC magnetic field in 
the shear plane ($B_{\rm y}$ with $E_{\rm z}$), perpendicular to the relative velocity in $x-$direction, while 
the $e^{\pm}$ cases generate AC electric and magnetic fields (Nishikawa et al. 2014b). 
These major findings obtained in the slab model, are found as well in this
new research along with new phenomena due to the cylindrical jet structure. The slab model was extended to the cylindrical mode (Nishikawa et al. 20014c), and also 
to the new model presented here, where a cylindrical jet is injected into the ambient  plasma.

Previously shock and velocity shear processes were investigated separately (e.g., Nishikawa et al. 2009a (shock), Nishikawa et al. 2014b (velocity shear)).  In order to  begin an investigation of the combined processes we performed simulations where a relativistic cylindrical jet is  injected  into ambient plasma, resulting in velocity shear and shocks in a potentially complicated shock/shear system as shown in Fig. \ref{fig2}c (global jet setup).  The present simulations utilize an injection method similar to that used in our previous studies of shocks (Nishikawa et al. 2009a;  Choi et al. 2014; Ardaneh et al. 2015) and includes a velocity shear between jet and ambient plasmas. Earlier simulations with similar setups were performed with significantly smaller jet radius and length (Nishikawa et al. 2003, 2005; Ng \& Noble 2006).  The simulations use plasmas with electron-proton ($e^{-}-p^{+}$ with $m_{\rm p}/m_{\rm e} = 1836$)  and electron-positron ($e^{\pm}$) compositions, so we can investigate the combination of kKHI, MI, and filamentation Weibel-like instabilities as a function of composition. 

This paper is structured as follows: Section 2 provides a summary of the theoretical analysis of kKHI and MI growth rates. Simulation setups illustrating the different study cases of the global jet simulations are described in Section 3.1 Simulation results of the cylindrical jet velocity shear and the global jet propagation, non-linear evolution and electromagnetic fields and currents structure are presented and discussed in Sections 3.2 and 3.3, respectively. Our findings are summarized in Section 4, where their application to AGNs and GRBs are discussed.

\vspace{-0.7cm}
\section{Theoretical Analysis Summary of kKHI and MI Growth Rates}
\vspace{-0.2cm}

Growth rates of the kKHI have been investigated theoretically, e.g., Nishikawa et al. (2014b), by analyzing the stability of longitudinal electrostatic perturbations considering a sharp velocity shear surface at $z = 0$ with ``jet'' plasma ($n_{\rm jt}$) at $z > 0$ and ``ambient'' plasma ($n_{\rm am}$) at $z < 0$ with the flow in jet and ambient plasma in  the $x-$direction. It is assumed that density, velocity, current and electric field perturbations are along the flow ($x-$axis) and of the form
\begin{equation}
f_{1}(x, z, t ) = f_{1}(z)e^{i({\bf k}x-\omega t)}, 
\end{equation}
where the wavevector ${\bf k} \equiv k_{\rm x}$ is parallel to the flow direction. We are considering a velocity shear surface that is infinite and transverse to the flow direction and perturbations are independent of $y$, i.e., $k_{\rm y} = 0$. The perturbed magnetic field lies along the $y-$axis, transverse to the flow and parallel to the shear surface. In the low-wavenumber limit 
($kc\ll \omega_{p}$)
with $v_{\rm am} = 0$ and $\gamma_{\rm am} = 1$, relevant to the numerical simulations, an analytic solution to the dispersion relation is given by
\begin{equation}
\omega  \sim \frac{(\gamma_{\rm jt}\omega_{\rm p,am}/\omega_{\rm p,jt})}
{(1 + \gamma_{\rm jt}\omega_{\rm p,am}/\omega_{\rm p,jt})}
kv_{\rm jt} \pm i\frac{(\gamma_{\rm jt}\omega_{\rm p,am}/\omega_{\rm p,jt})^{1/2}}
{(1 + \gamma_{\rm jt}\omega_{\rm p,am}/\omega_{\rm p,jt})}
kv_{\rm jt}~, 
\end{equation}
where 
$$
 \omega_{p}^2 \equiv {4 \pi n_e e^2 \over \gamma^3 m_e}~.
$$
In previous simulations the phase (drift) velocity was comparable to the jet speed and the low-wavenumber growth rate scaled with $\gamma^{-5/4}_{\rm jt}$ (Nishikawa et al. 2014b). Numerical solution of the dispersion suggested that
\begin{equation}
\omega_I^* \sim  \omega_{p,jt} ~~{\rm and}~~  k^*v_{jt} \sim {(1 +  \gamma_{jt}\omega_{p,am}/\omega_{p,jt})\over ( \gamma_{jt}\omega_{p,am}/\omega_{p,jt})^{1/2}} \omega_{p,jt}, 
\end{equation}
provided zeroth order estimates for the maximum temporal growth rate, $\omega_I^* \propto \gamma^{-3/2}$,  and wavenumber, $k^*$, at maximum growth.

Alves et al. (2015) analyzed the stability of electromagnetic perturbations in the transverse plane of a collisionless plasma sheared flow with velocity profile $v_{0} = v_{0}(x)e_{\rm z}$. 
Here we adapt their numerical analysis to our present simulations of 3D global jets and assume the slab geometry used in previous work (Nishikawa et al. 2014b). Now the perturbations are of the form
\begin{equation}
f_{1}(x, z, t ) = f_{1}(z)e^{i({\bf k}y-\omega t)}, 
\end{equation}
and the wavevector ${\bf k} \equiv k_{\rm y}$ is transverse to the flow direction. Based on the analysis by Alves et al. (2015) we assume  a slab setup as in their case and our previous cases and ignore the effects of  ambient plasma in the jet.
In the present case, $n_{\rm jt} = n_{\rm am} = n_{0}$ (for simplicity) and $v^{+} = v_{\rm jt}, \, v^{-} = v_{\rm am}= 0$, the growth rate becomes
\begin{equation}
\frac{\Gamma}{\omega_{\rm pe}}= \frac{1}{\sqrt{2}}\left( \sqrt{\frac{4k^{2}_{\rm y}v_{\rm jt}^{2}}{\gamma_{\rm jt}\omega^{2}_{\rm pe}}+ D^{2}_{\parallel}}- D_{\parallel}\right )^{1/2},
\end{equation}
where $\Gamma= {\rm Im}(\omega)$ and $D_{\parallel} = 1/\gamma_{\rm jt}^{3}+ k^{2}_{\rm y}c^{2}/\omega^{2}_{\rm pe}$. The fastest growth of this unstable branch ($\partial_{\rm k}\Gamma = 0$)
is found at $k_{\rm y}\rightarrow \infty$ (finite thermal effects and/or smooth velocity shear profiles introduce a cutoff at finite $k$).  In the limit, $n_{\rm am} \gg n_{\rm jt}/\gamma^{3}_{\rm jt}, \, \gamma_{\rm jet} \gg 1$, and $v_{\rm am}= 0$ the cutoff at finite $k$ provides a maximum growth rate, $\Gamma_{\max}$, for a given sheared flow Lorentz factor of $\Gamma_{\max}/\omega_{\rm pe} = v_{\rm jt}/c
\sqrt{\gamma_{\rm jt}}$. 

The MI and ESKHI (Alves et al. 2012) growth rates are compared for
different shear Lorentz factors and velocities in Figure 1 of Alves et al. (2015). While the ESKHI has higher growth rate than the MI for non-relativistic flows, for relativistic flows the MI growth rate
declines as $\gamma_{\rm jt}^{-1/2}$, and declines slower than the ESKHI, which declines as $\gamma_{\rm jt}^{-3/2}$ (see Fig. 1b in Alves et al. 2015). Provided that initial perturbations for both instabilities are similar, the MI is the dominant electron-scale instability in relativistic shear flows.

The application of the analytical description of kKHI and MI, and comparison with the results 
of this simulation study, is limited because our simulation setup does not use  a slab model, and the jet density is 
added to the ambient density, that is not included in the analytical description.  Furthermore, the expression for the growth rate in the large wavelength limit with the same densities for jet and ambient plasmas is not applicable to the simulation results. However, the basic qualitative results in Alves et al. (2015) would be applied to our simulations 
which are described in the later sections with the fact that the MI mode is dominant in the jets with high Lorentz factor. The growing kKHI and MI are observed in jet  structures. The kKHI and MI are longitudinal and transverse modes, respectively. Therefore, the excited modes of kKHI modes are found in the $x-z$ plane, and, on the contrary, the MI mode is found in  the $y-z$ plane as illustrated in Alves et al. (2016). In 3D displays both modes are seen simultaneously.

\vspace{-0.7cm}
\section{Global Jet Simulations of kKHI, MI and Weibel-like Instabilities}
\vspace{-0.2cm}

\subsection{Previous and Present Simulation Setups}
\vspace{-0.2cm}

\begin{figure}[ht]
\epsscale{0.33}
\plotone{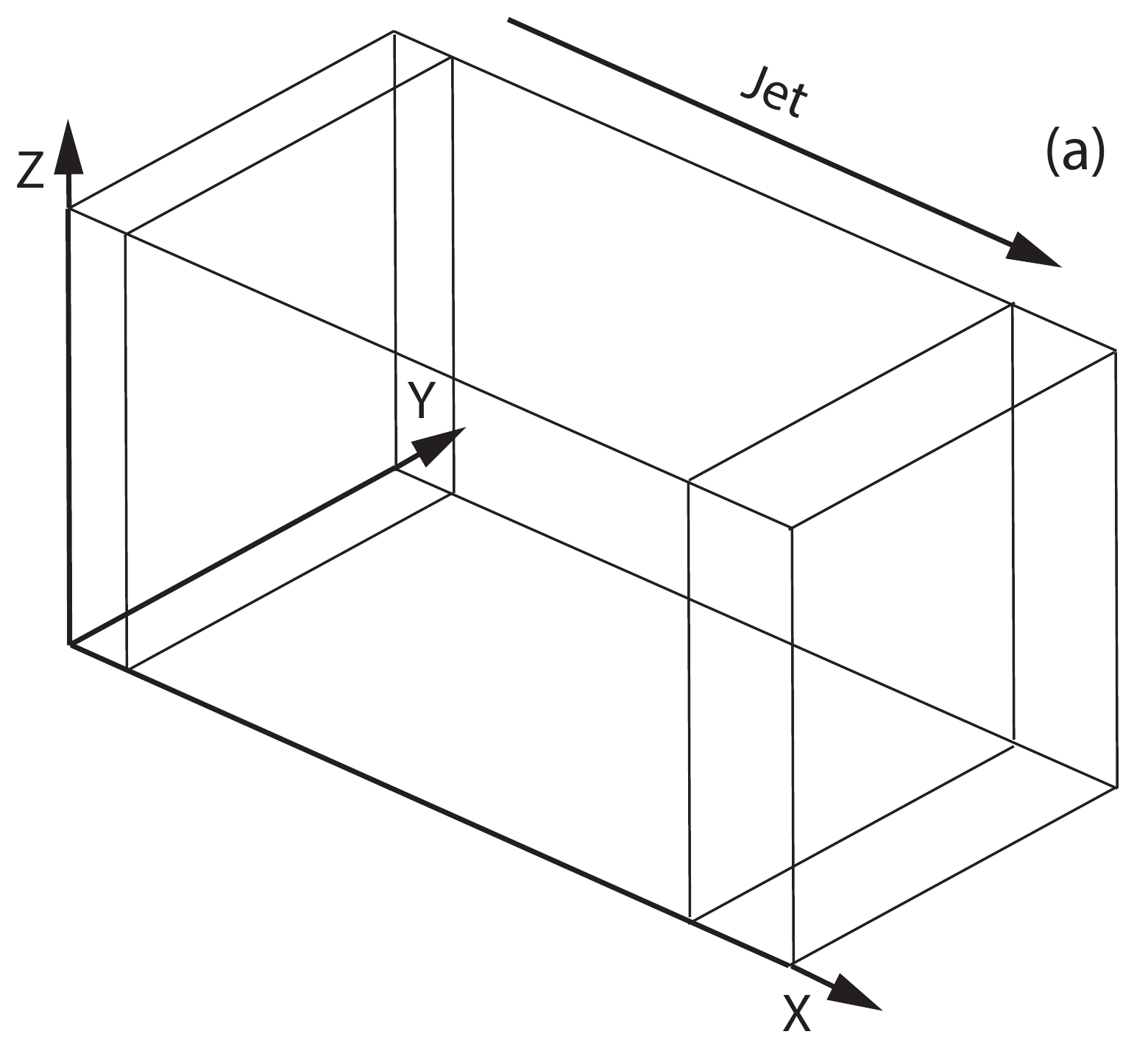}
\epsscale{0.28}
\hspace*{-0.3cm}
\plotone{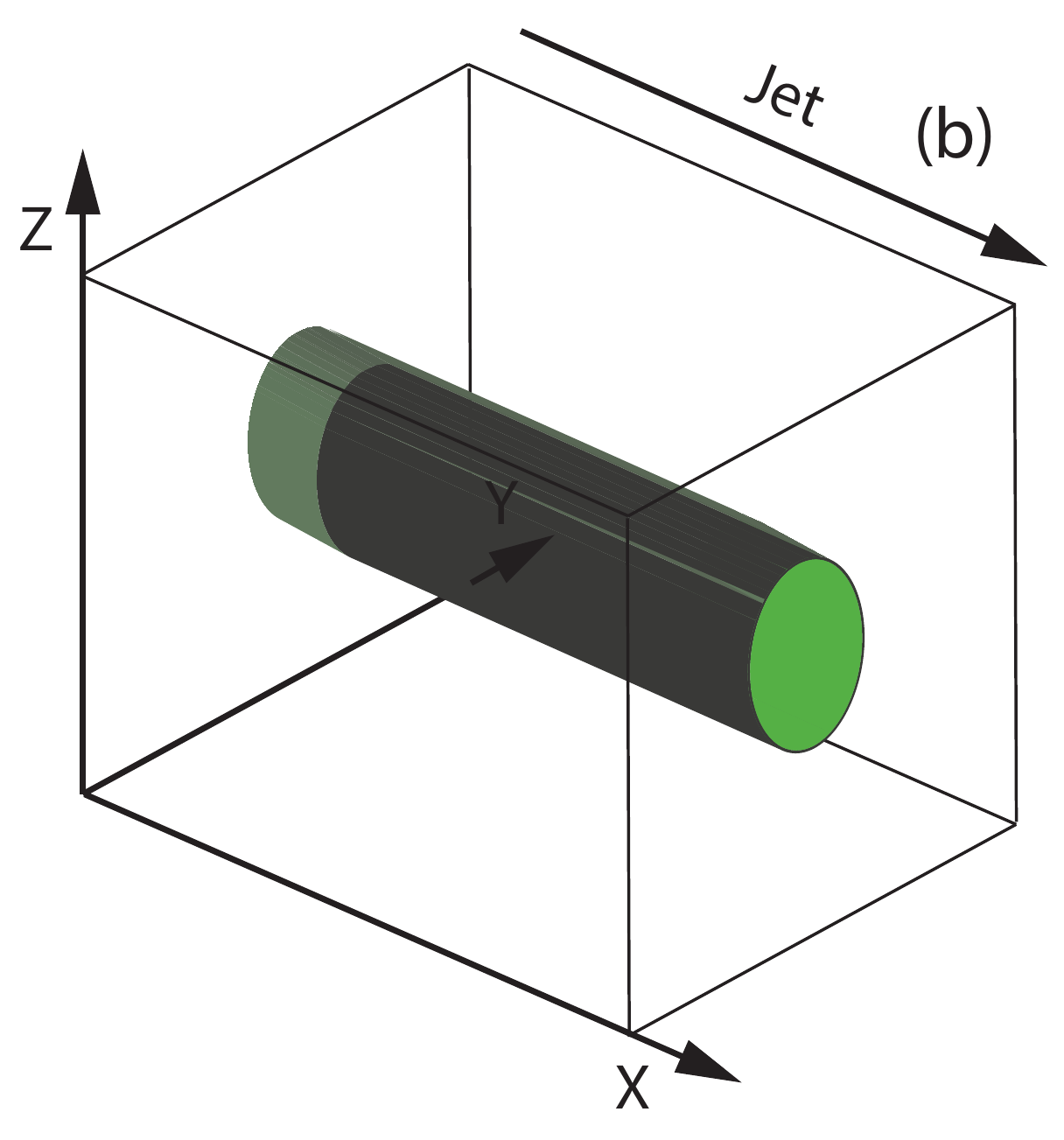}
\epsscale{.33}
\hspace*{-0.3cm}
\plotone{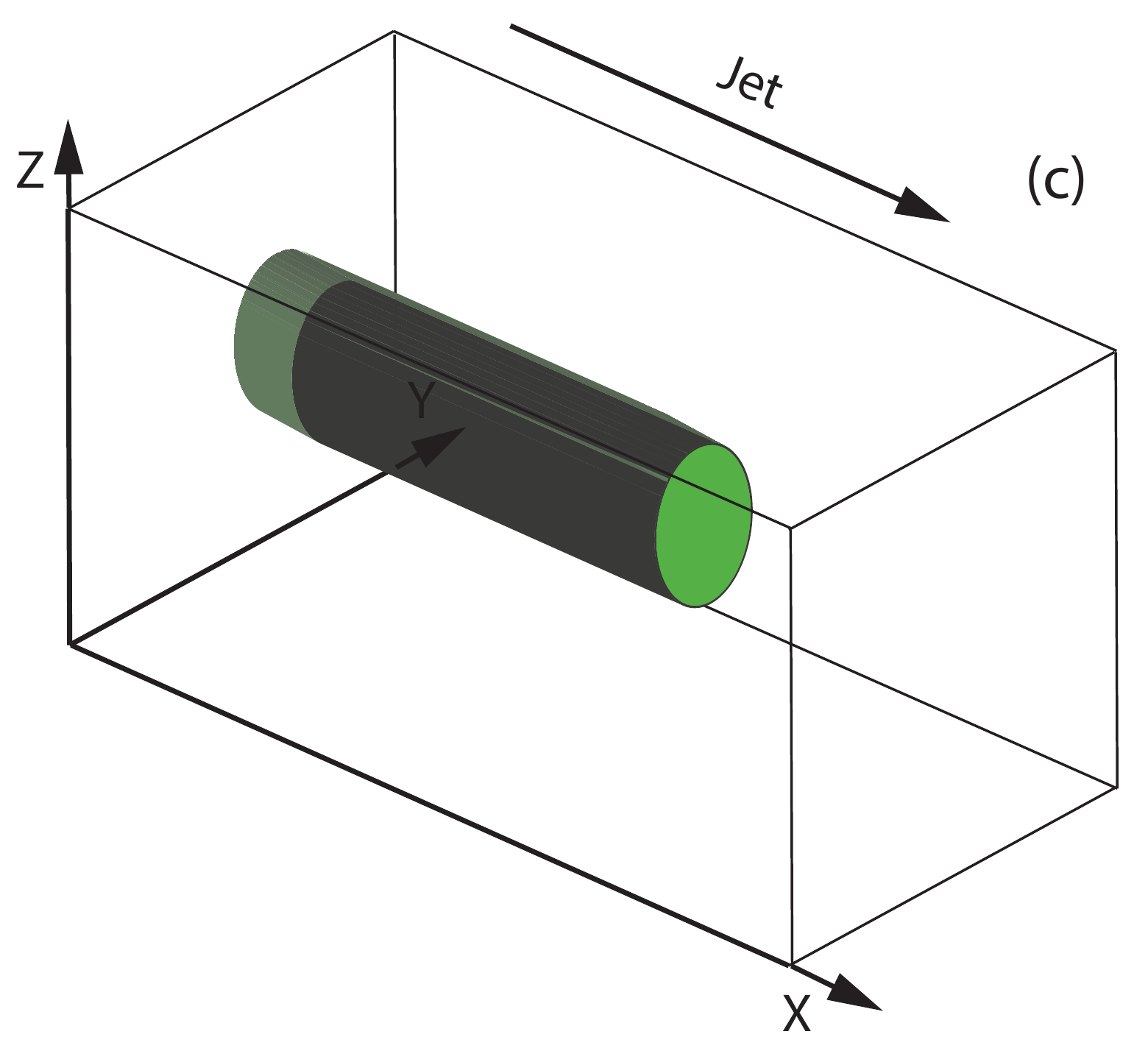}
\vspace*{-0.2cm}
\caption{\footnotesize \baselineskip 11pt Schematic simulation setups: (a) injection scheme for shock simulations where jets are injected at $x= 25\Delta$ in the $y - z$ plane (e.g., Nishikawa et al. 2009a), (b) cylindrical injection scheme for shear flow simulations where jets are initially placed along the entire length of the $x-$axis at the center of the $y - z$ plane (Alves 2010; Nishikawa et al. 2014c), and (c) global jet injection scheme (for this simulation study) where the jet is injected at $x = 100\Delta$ with jet radius $r_{\rm jt} = 100\Delta$ at the center of the $y - z$ plane (not scaled). \label{fig2}}
\end{figure}

We perform ``global'' simulations of the injection of a cylindrical jet into an ambient plasma in order to investigate shocks (Weibel-like instabilities) and velocity shear (kKHI and MI) simultaneously. 
Simulations of shocks generated by the Weibel instability have previously been performed using the model shown in Figure \ref{fig2}a where the jet spans the computational grid in the transverse direction (e.g., Nishikawa et al. 2009a; Choi et al. 2014; Ardaneh et al. 2015). In other simulations, the  kKHI and MI have been investigated using a slab jet model (not shown) (e.g., Nishikawa et al.  2013, 2014a,b,c), or a   cylindrical model where the jet spans the computational grid in the longitudinal direction (Alves 2010; Nishikawa et al. 2014c) as shown in Figure \ref{fig2}b. 

\vspace{-0.10cm}
The separate investigation of leading edge shocks or velocity shear surfaces does not provide a realistic scenario for jets in most astrophysical systems.
To achieve and investigate the combined effect of the leading edge shocks and velocity shear instabilities, such as might be associated with the ``needles-in-a-jet" or ``jet-in-a-jet" scenarios proposed in the blazar AGN context,  we perform simulations of a relativistic cylindrical jet injected into an ambient plasma.  Such simulations contain both a velocity shear and a shock in a potentially  complicated shock/shear system, as shown in Figure \ref{fig2}c.  We use an open boundary condition for the bottom and top (at $x = x_{\min}, x_{\max}$). 
In order to avoid possible artificial phenomena which may affect the initial jet evolution, we inject the jet far from the boundary at $x = 100\Delta$. The jet is injected with a sharp edge and the ``top hat'' shape. It should be noted that in the case (c) the jet is injected into the ambient plasmas therefore there is no discontinuity as in the case (b) in which the jet is set inside the ambient plasma, with a discontinuity between the two plasmas. 
However, the jet is injected with a sharp edge, therefore there exists some discontinuity initially
at the jet boundary which excites kKHI and MI. Inside the jet the Weibel instability is excited interacting with
the ambient plasma.

\vspace{-0.7cm}
\subsection{Cylindrical Jet Velocity Shear Result Summary}
\vspace{-0.2cm}

In the following we describe kKHI and MI simulation results from a cylindrical velocity shear setup as shown in Figure \ref{fig2}b, in order to illustrate the basic  differences between  $e^{\pm}$ and $e^{-}-p^{+}$  plasma jets (Nishikawa et al. 2014c). These simulations have been performed using a numerical grid with $(L_{\rm x}, L_{\rm y}, L_{\rm z}) = (1005\Delta, 205\Delta, 205\Delta)$ (simulation  cell size: $\Delta = 1$) and periodic boundary conditions in all dimensions. The jet and sheath (electron) plasma number density measured in the simulation frame is $n_{\rm jt}= n_{\rm am} = 8$.  The cylindrical jet with jet radius $r_{\rm jt} =50\Delta$ is inserted in the middle of the $y-z$ plane. The electron skin depth, $\lambda_{\rm s} = c/\omega_{\rm pe} = 12.2\Delta$, where $\omega_{\rm pe} = (e^{2}n_{\rm am}/\epsilon_0 m_{\rm e})^{1/2}$ is the electron plasma frequency and the electron Debye length for the ambient electrons $\lambda_{\rm D}=1.2\Delta$.  The jet-electron thermal velocity is $v_{\rm jt,th,e} = 0.014c$ in the jet reference frame, where $c$ is the speed of light.  The electron thermal velocity in the ambient plasma is $v_{\rm am,th,e} = 0.03c$, and ion thermal velocities are smaller by $(m_{\rm i}/m_{\rm e})^{1/2}$. Simulations were performed using an electron-positron ($e^{\pm}$) plasma or an electron-proton ($e^{-}-p^{+}$ with $m_{\rm p}/m_{\rm e} = 1836$) plasma for the jet Lorentz factor of 5 and with the sheath plasma at rest ($v_{\rm sheath}= 0$).

\begin{figure}[h!]
\hspace{1.4cm}
\includegraphics[width=65mm]{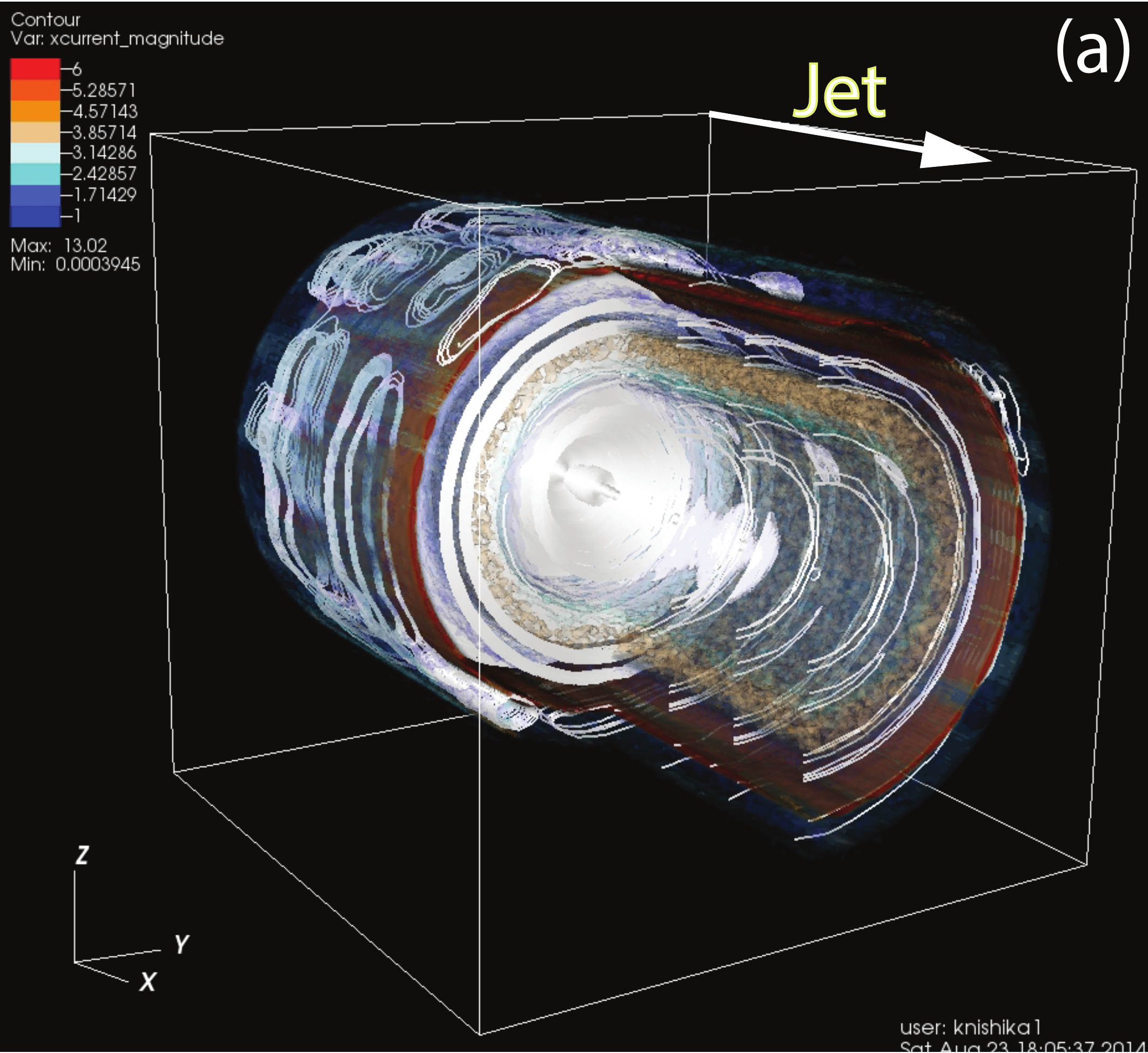}
\includegraphics[width=65mm]{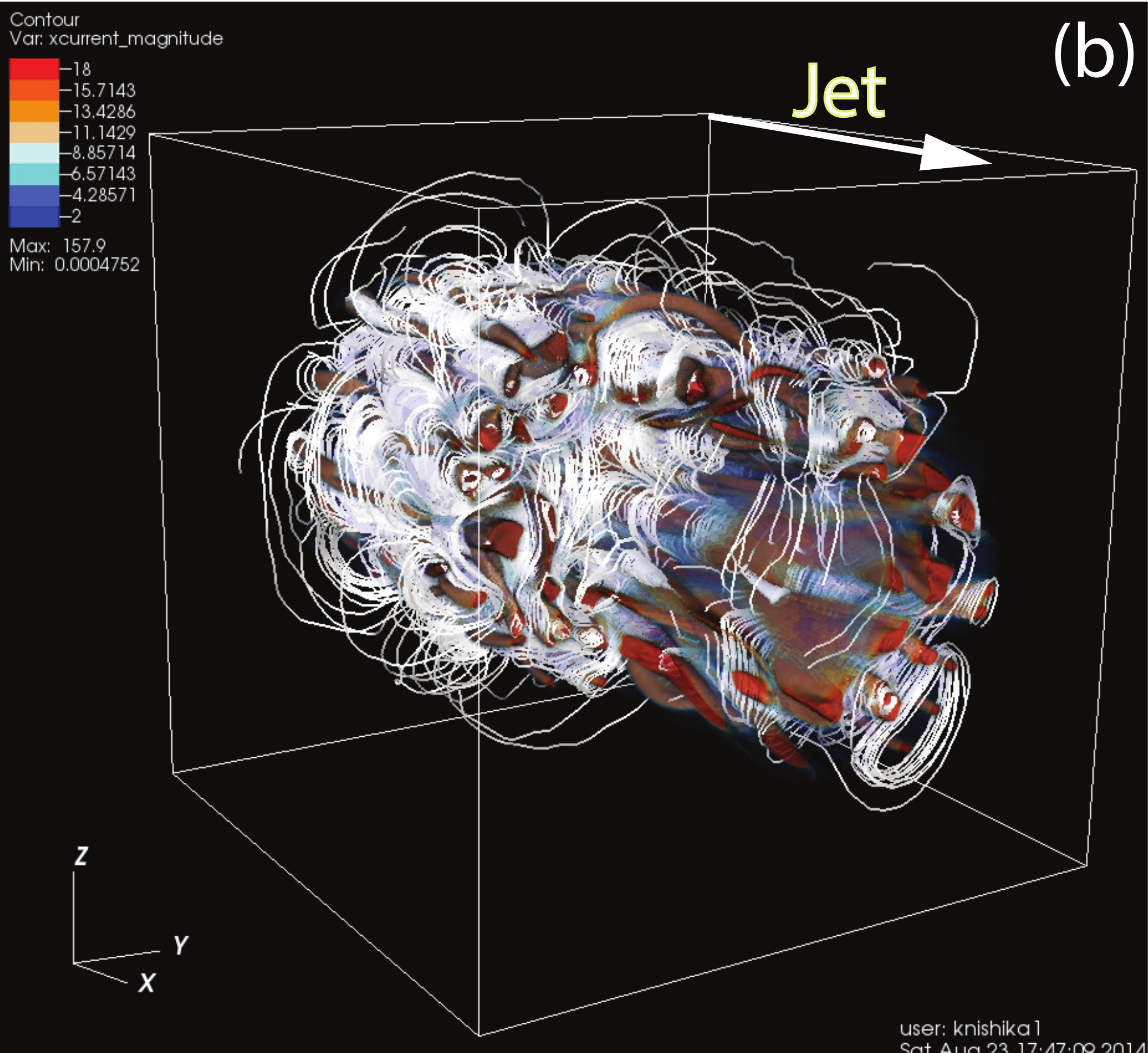}
\vspace{-0.cm}
\caption{\footnotesize \baselineskip 11pt {Isocontour plots of the $J_{\rm x}$ current magnitude with magnetic field lines (one fifth of the jet length)  for (a) an $e^{-}-p^{+}$ and (b) an $e^{\pm}$ jet  at simulation time $t =  300\omega_{\rm pe}^{-1}$. The 3D displays are clipped along and perpendicular to the jet 
in order to view the interior. Color bars (a) 1.0 - 6.0; (b) 2 - 18}}
\label{fig1}
\vspace{-0.0cm}
\end{figure}

Figure \ref{fig1} shows isocontour images of the $x$ component of the current along  with magnetic field lines generated by the kKHI and MI for $e^{\pm}$ and $e^{-}-p^{+}$  jets.
\vspace*{-0.0cm}
In the  $e^{-}-p^{+}$ jet case shown in Figure \ref{fig1}a, currents are generated in sheet like layers and magnetic fields are wrapped around the jet. The toroidal magnetic field lines outside of the jet show signatures of both kKHI and MI (Fig. \ref{fig1}a).  Nishikawa et al. (2014b) have shown the development of the MI mode at the earlier linear stage even they had not yet identified it as an MI. At the nonlinear stage the wavelength of MI becomes very large and DC magnetic field is generated as shown in Alves et al. (2012). In this simulation the cylindrical jet is used, the DC magnetic field becomes more eminent and related to the collimation of jet. As it is discussed in Section 2 for relativistic jets with higher Lorentz factor the MI becomes dominant compared to kKHI (Alves et al. 2015), and we believe that the MI is more likely the source for generating the DC magnetic field.   On the other hand, in the  $e^{\pm}$ jet case shown in Figure \ref{fig1}b, many distinct current filaments are generated near the velocity shear, and the individual current filaments  are wrapped by the magnetic field which clearly indicates the MI. The clear difference 
in the magnetic field structure between these two cases, may make it possible to distinguish different jet 
compositions via differences in circular and linear polarization.  

\vspace{-0.7cm}
\subsection{Results of 3D Global Jet Simulations}
\vspace{-0.2cm}

Nishikawa et al. (2014c) reported simulation results using a computation system with $(L_{\rm x}, L_{\rm y}, L_{\rm z})  = (645\Delta,   131\Delta, 131\Delta$), jet Lorentz factor $\gamma_{\rm jt} = 10$, jet injection at $x=100\Delta$ and jet radius $r_{\rm jt} =  20\Delta$ in the center of the $y-z$ plane. However, the jet radius in these simulations was too small to clearly distinguish velocity shear and shock effects.  Here we overcome this deficiency by using a 5 times larger jet radius, $r_{\rm jt} =  100\Delta$, and a suitably larger computation system with $(L_{\rm x}, L_{\rm y}, L_{\rm z}) = (2005\Delta,  1005\Delta, 1005\Delta$).
The cylindrical jet with jet Lorentz factor $\gamma_{\rm jt} = 15$ is continuously injected at $x = 100\Delta$ in the middle of the $y-z$ plane as shown in Figure \ref{fig2}c. Note that this system is short compared to previous pure shock simulations (Nishikawa et al. 2009a; Choi et al. 2014; Ardadeh et al. 2015) and reveals only the earliest stages of shock development.

The jet and ambient (electron) plasma number density measured in the simulation frame is $n_{\rm jt}= 8$
and $n_{\rm am} = 11.1$, respectively. The electron skin depth in the ambient medium is $\lambda_{\rm s} = 10.4\Delta$, and the electron Debye length for the ambient electrons is $\lambda_{\rm D} =  1.2\Delta$.  The jet-electron thermal velocity is $v_{\rm jt,th,e} = 0.014c$ in the jet reference frame.  The electron thermal velocity in the ambient plasma is $v_{\rm am,th,e} = v_{\rm am,th,p}= 0.05c$, and ion thermal velocities are smaller by $(m_{\rm i}/m_{\rm e})^{1/2}$. 
We have chosen these thermal velocities in order to investigate the combined effect of shocks (e.g., Nishikawa et al. 2009a) and kKHI (e.g., Nishikawa et al. 2014b) with thermal temperatures similar to those used in these studies.
Simulations were performed using an electron-positron plasma or an electron-proton  plasma with the ambient plasma at rest ($v_{\rm am}= 0$). The system evolution is followed until $t=1700 \omega_{\rm pe}^{-1}$.

\begin{figure}[h!]
\begin{center}
\epsscale{0.9}
\vspace*{-0.3cm}
\plotone{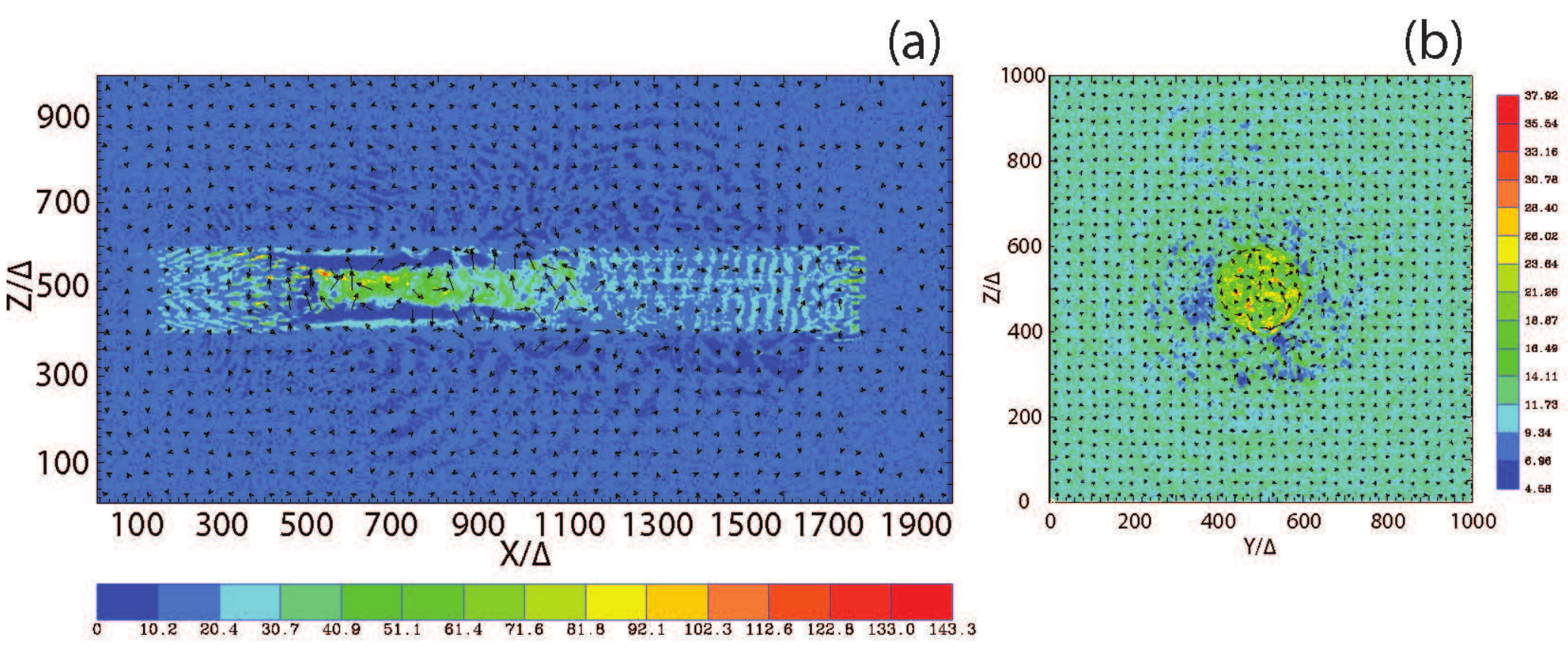} 
\vspace*{-0.7cm}
\epsscale{0.9}
\plotone{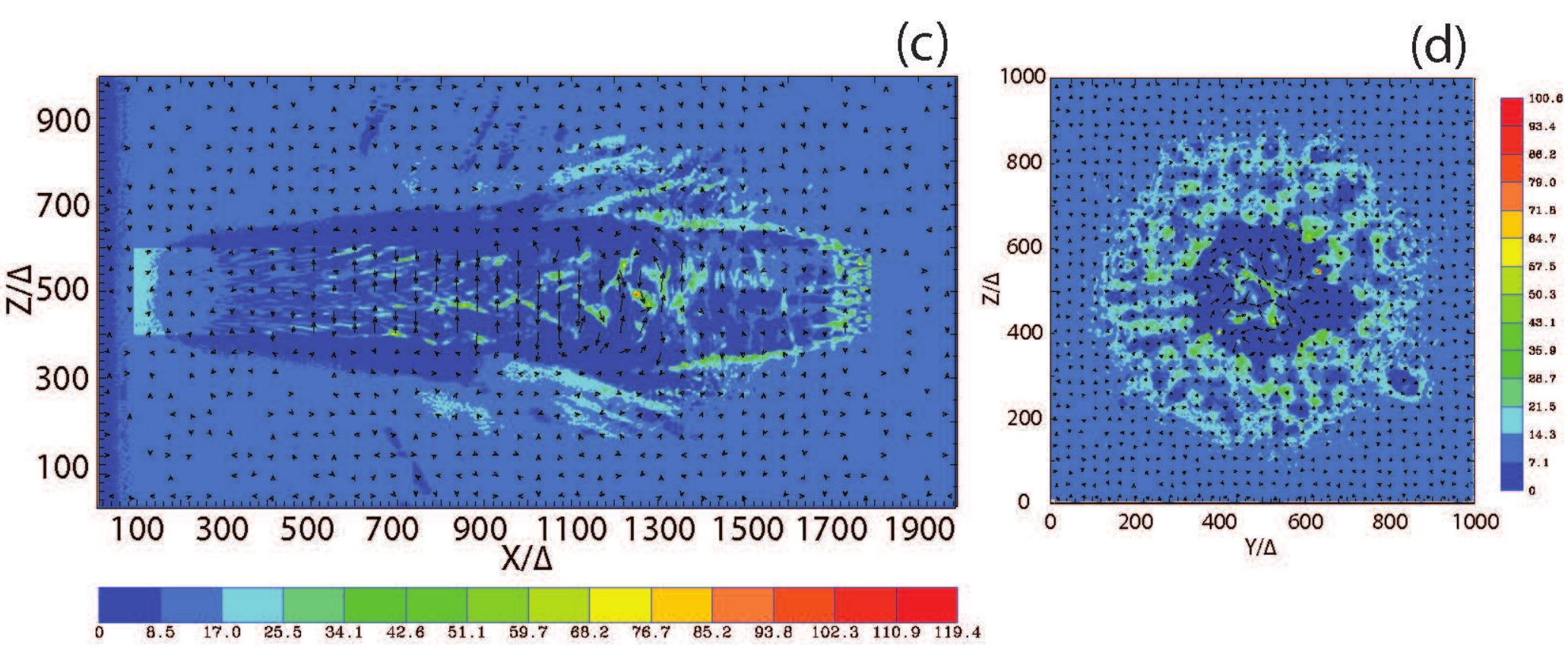}
\end{center}
\vspace*{-0.3cm}
\caption{\footnotesize \baselineskip 11pt  {Electron density with magnetic field arrows in the plane at $t=1700 \omega_{\rm pe}^{-1}$ for (upper panels) the $e^{-}-p^{+}$ case and (lower panels) the  $e^{\pm}$ case.
Panels (a) and (c) show electron density in the $x-z$ plane at $y=500\Delta$ and panels (b) and (d) show electron density in the $y-z$ plane at $x=1200\Delta$, where development is in the nonlinear stage. Color bars (a) 0 - 143.3;
(b) 4.58 - 37.92; (c) 0 - 119.4; (d) 0 - 100.6}}   
\label{fig3}
\end{figure}

The global evolution of relativistic electron-positron and electron-proton plasma jets is shown in Figure \ref{fig3}. 
For the $e^{-}-p^{+}$ case jet collimation takes place at $500\Delta < x < 750\Delta$ due to the toroidal magnetic fields generated by the kKHI and the MI.  Jet electrons are pinched towards
the jet center where current filaments merge and high electron density is created beyond $x\approx 500\Delta$.
The collimation disappears gradually beyond $x\approx1000\Delta$.  The jet is well defined by the heavy jet protons. The wave patterns generated by the kKHI and the MI and extending for $r\gg r_{\rm jt}$
are shown in Figures \ref{fig3}a and  \ref{fig3}b.
For the $e^{\pm}$ case, since electrons and positrons are light, there is  a mix of  jet and 
ambient particles at the velocity shear boundary as found previously using the cylindrical core sheath scheme (Nishikawa et al. 2014a,b,c). As in shock simulations (Nishikawa et al. 2009a), the Weibel instability is excited in the jet and density fluctuation patterns are generated around $x\approx1250\Delta$. At the same time the kKHI and  the MI are excited in the velocity shear region causing jet and ambient electrons to move away from the jet boundary. This motion appears as the streaks in Figure \ref{fig3}c. The excited wave patterns are 
clearly exhibited in Figure \ref{fig3}d.

\begin{figure}[h!]
\begin{center}
\epsscale{0.9}
\vspace*{-0.3cm}
\plotone{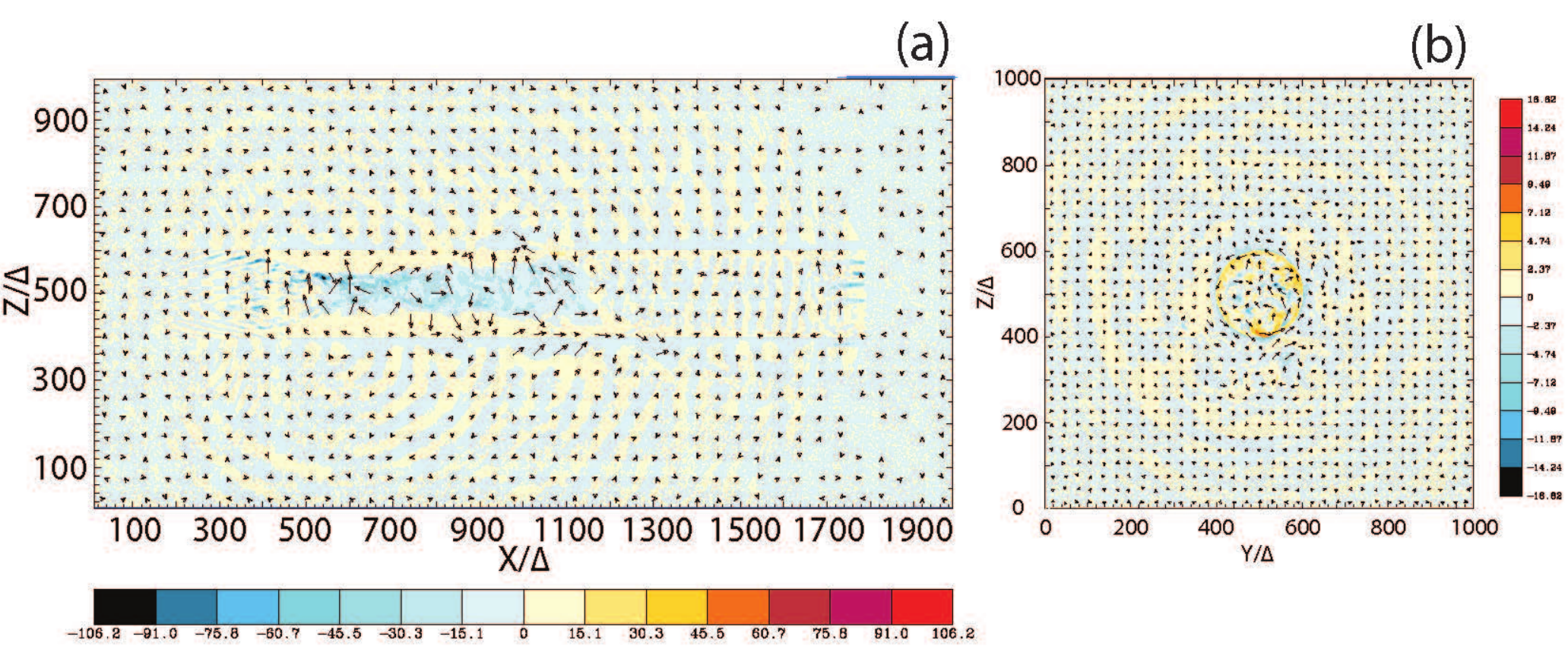} 
\vspace*{-0.3cm}
\epsscale{0.9}
\plotone{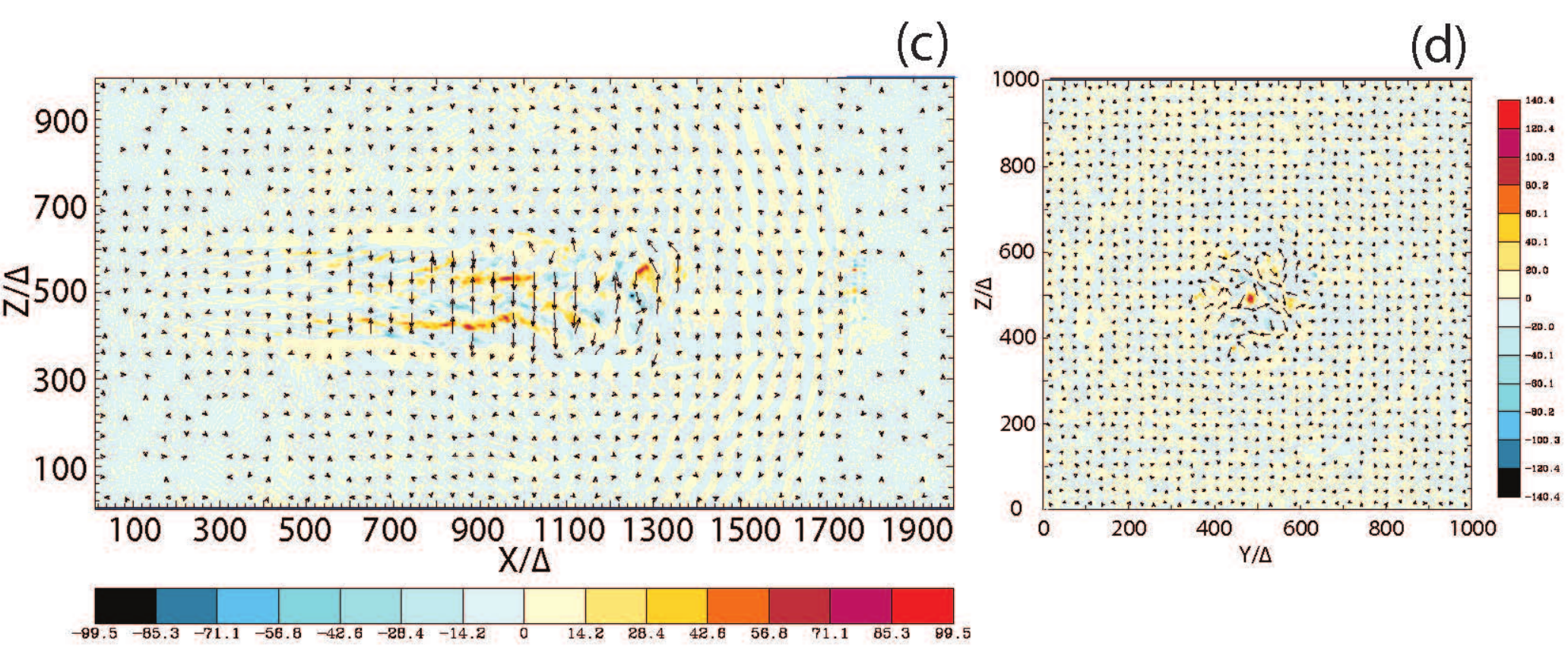}
\end{center}
\vspace*{-0.5cm}
\caption{\footnotesize \baselineskip 11pt The $x$ component of the electron current density $J_{\rm x}$ with magnetic field arrows in the plane at $t=1700 \omega_{\rm pe}^{-1}$ for (upper panels) the $e^{-}-p^{+}$ case and (lower panels) the $e^{\pm}$ case. Panels (a) and (c) show the electron density in the $x-z$ plane at $y=500\Delta$ and panels (b) and (d) show a cross section in the $y-z$ plane at $x=1200\Delta$. Color bars (a) $\pm 106.2$; (b) $\pm 16.62$; 
(c)  $\pm 99.5$; (d) $\pm 140.4$ \label{fig4}}
\end{figure}

Differences between the two cases are also revealed in other physical quantities, such as the $x$ component of current density shown in  Figure \ref{fig4}. 
Magnetic fields in the plane are shown by lengthened arrows so that even weak magnetic fields can be seen. 
For the $e^{\pm}$ case the Weibel instability generates current filaments shown in Figure \ref{fig4}c
that nonlinearly saturate around $x\approx 1250\Delta$. In the nonlinear region electron
density fluctuations are a sign of shock formation. The initial nonlinear stage (minor saturation) is 
seen in the merging of current  filaments as also seen in the previous simulation in Nishikawa et al. (2009a).
Note, however, that in order to fully establish the shock we 
require a grid length that is larger than $4000\Delta$ (see, e.g., Nishikawa et al. 2009a).
For the $e^{-}-p^{+}$ case, strong electron currents are generated by the high electron density
in the collimation region shown in Figure \ref{fig4}a.  The jet boundary is
framed by the proton currents shown in Figures \ref{fig4}a and \ref{fig4}b. Wave patterns generated 
by the kKHI and MI are clearly seen as radial propagation in the plane perpendicular to the jet.

\begin{figure}[h!]
\begin{center}
\epsscale{1.1}
\vspace*{-0.2cm}

\vspace*{-0.05cm}
\plottwo{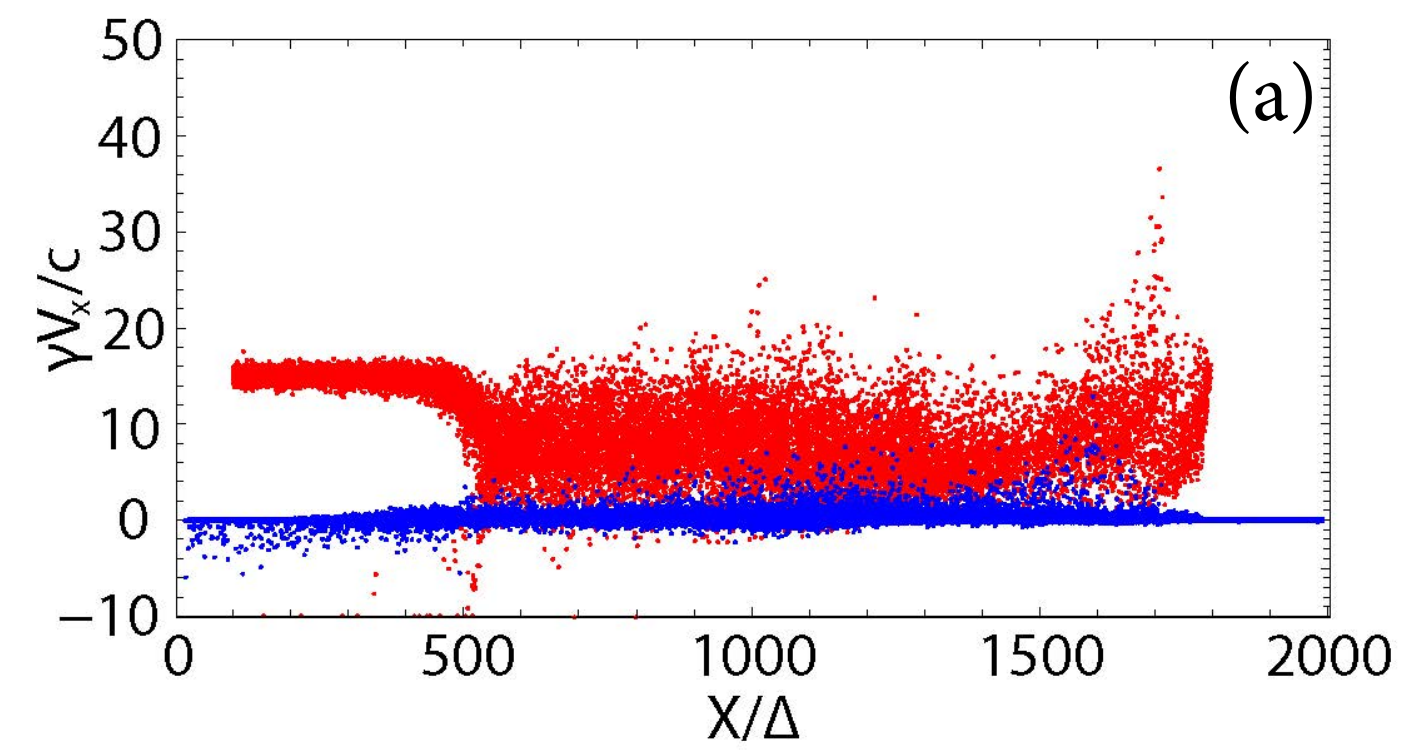}{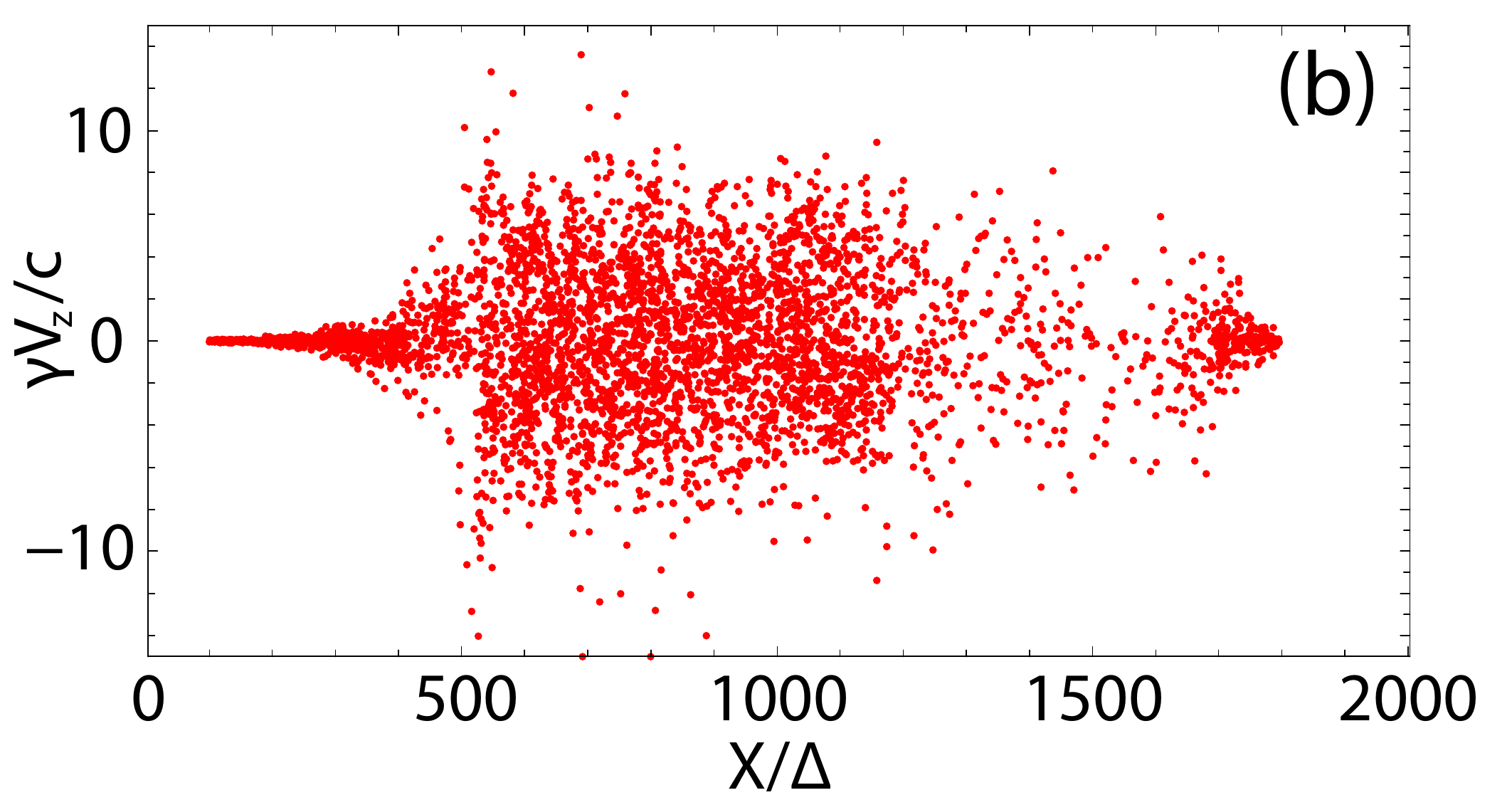} 
\vspace*{-0.1cm}

\vspace*{-0.05cm}
\plottwo{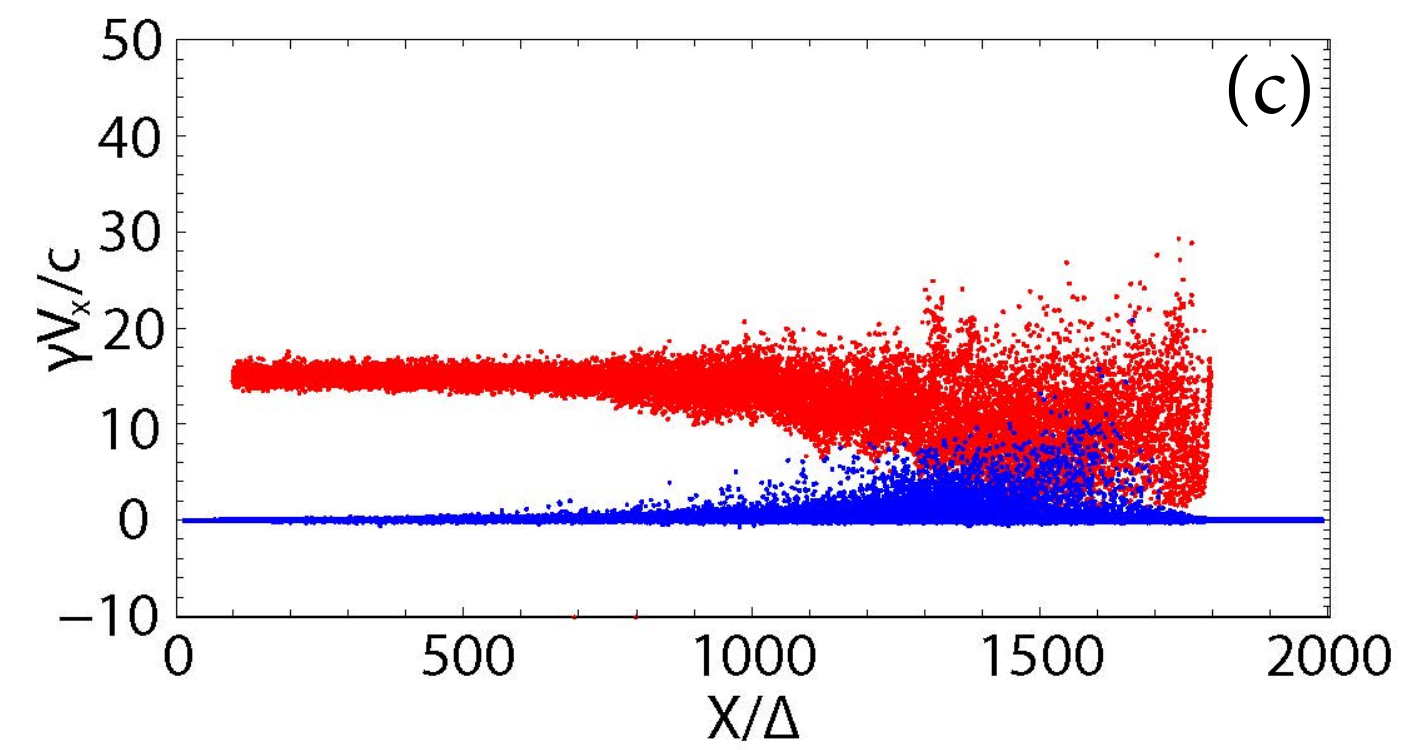}{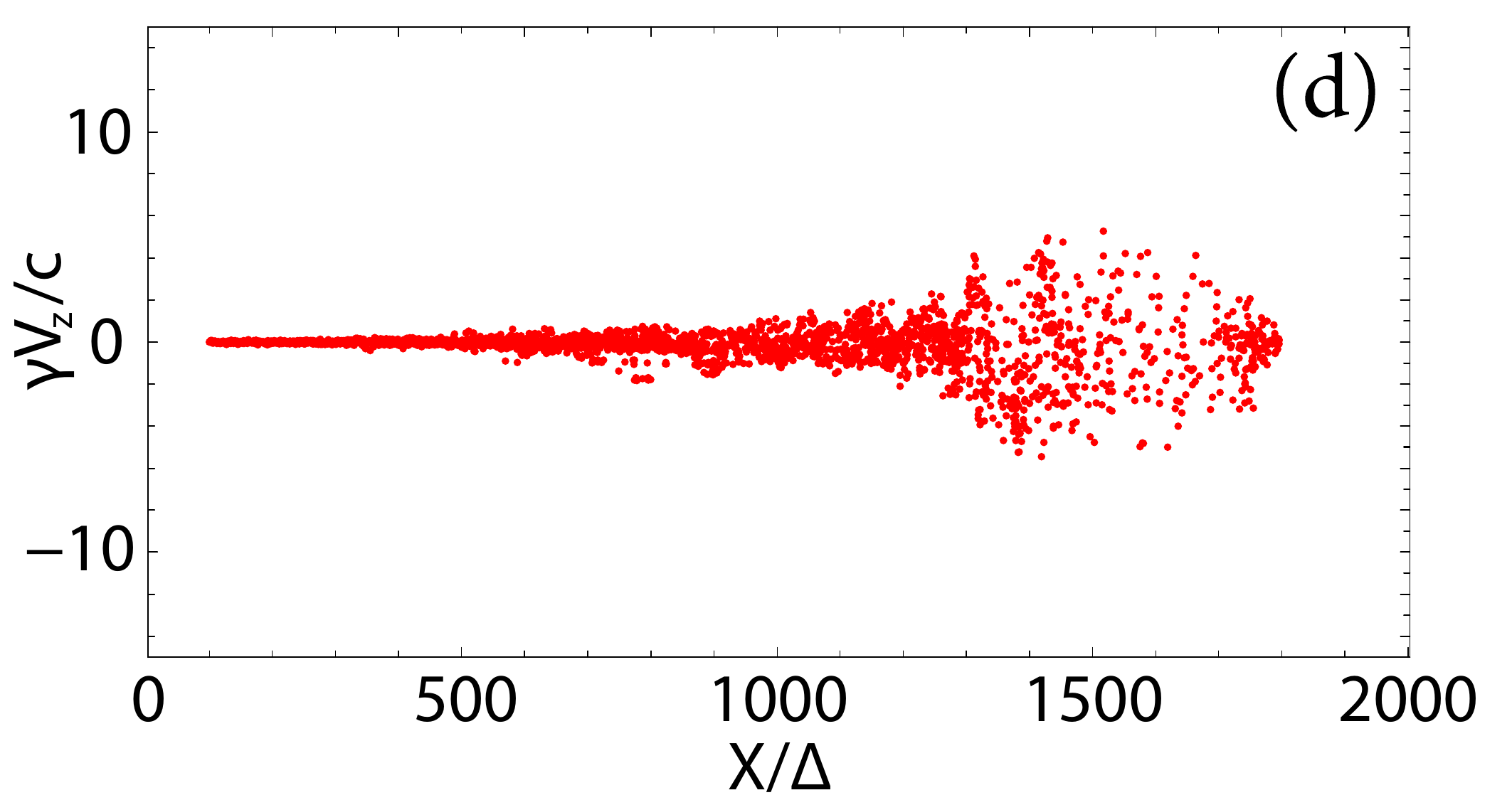} 
\end{center}
\vspace*{-0.6cm}
\caption{\footnotesize \baselineskip 11pt Phase-space distribution of jet (red) and ambient (blue) electrons at $t=1700 \omega_{\rm pe}^{-1}$ for (upper panels) the $e^{-}-p^{+}$ case and (lower panels) the $e^{\pm}$ case. Panels (a) and (c) show particle $x-\gamma v_{\rm x}$  phase-space, and panels (b) and (d) show  particle $x-\gamma v_{\rm z}$  phase-space.  \label{fig5}}
\end{figure}

Figure \ref{fig5} shows the particle $x-\gamma v_{\rm x}$  phase-space (panels (a) and (c)) and  the particle 
$x-\gamma v_{\rm z}$   phase-space (panels (b) and (d)) for jet electrons (red) and ambient electrons (blue). 
The location of jet collimation in the $e^{-}-p^{+}$ case is where jet electrons are strongly slowed 
and some jet electrons are reflected, i.e., the jet electrons have $\gamma v_{\rm x} <0$. At the collimation shock jet electrons are accelerated perpendicularly as shown in Figure \ref{fig5}b. Jet electrons are accelerated just behind the jet front as shown by the spike in phase space at $x\approx 1700\Delta$ in Figure \ref{fig5}a. Some ambient electrons are also slightly accelerated. The behavior of jet electrons is very similar to shock simulations (Nishikawa et al. 2009a) even though the region outside of the jet shows a complicated structure due to the excited kKHI and MI.
For the $e^{\pm}$ case, the jet and ambient electrons are slowly accelerated by the Weibel
instability. The ambient electrons are swept up as in the shock simulations (e.g., Nishikawa et al. 2009a; Choi et al. 1014; Ardaneh et al. 2015). However, due to the short simulation box the shock system has not yet fully formed. 

\begin{figure}[h!]
\begin{center}
\epsscale{0.9}
\vspace*{-0.2cm}
\plotone{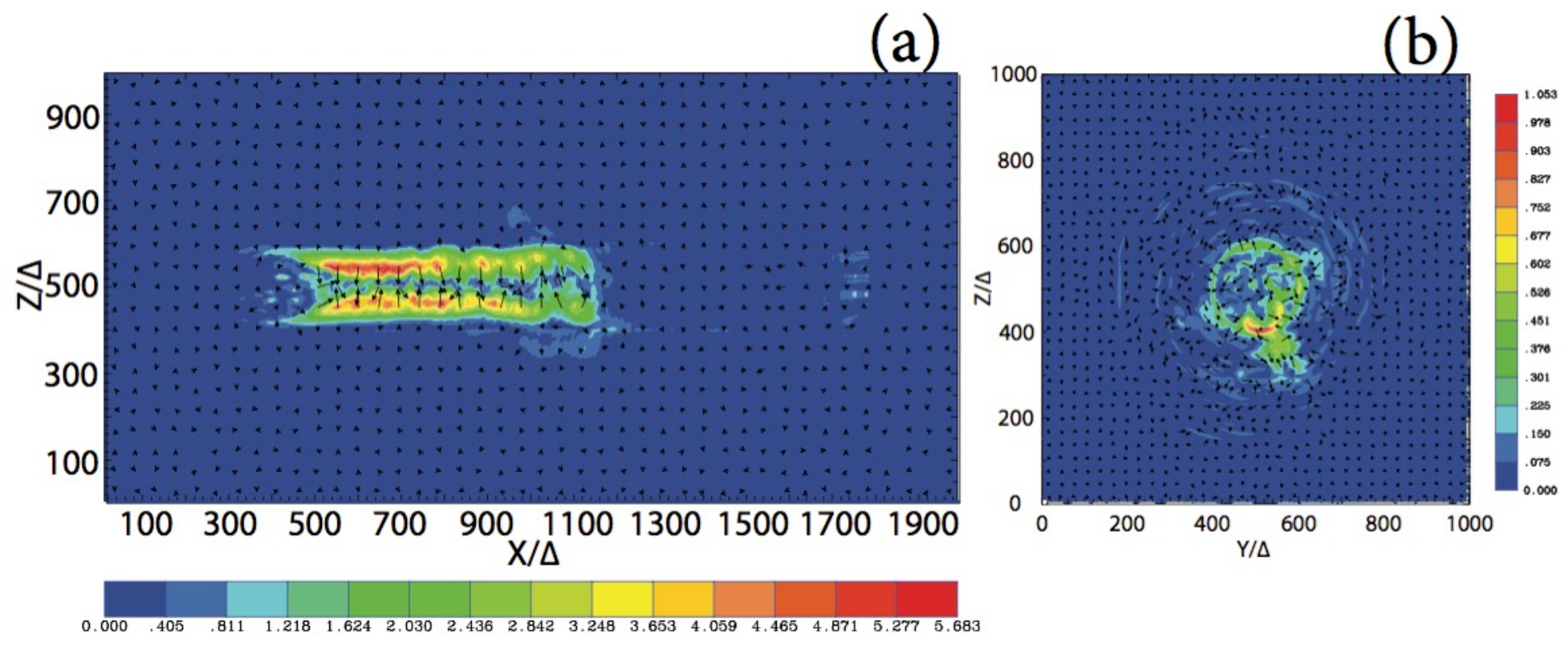} 
\vspace*{-0.2cm}
\epsscale{0.9}
\plotone{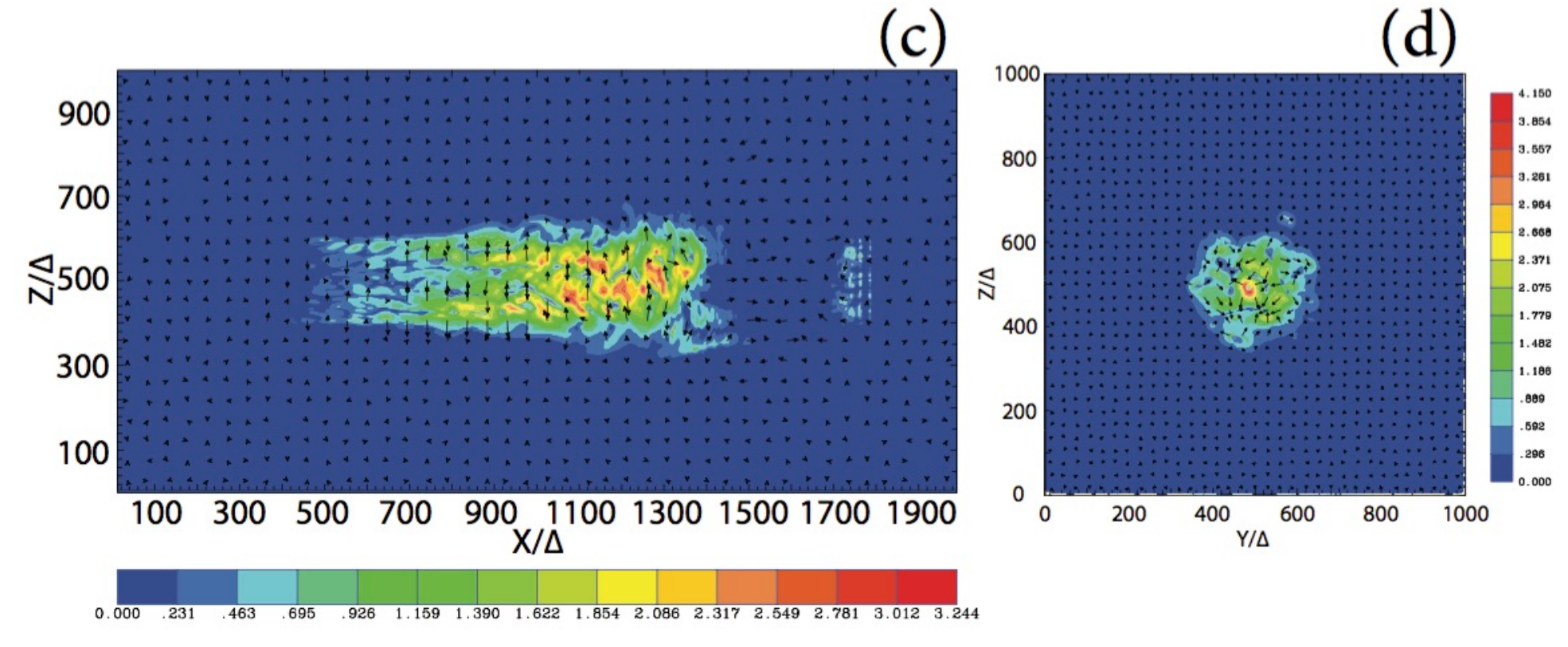}
\end{center}
\vspace*{-0.7cm}
\caption{\footnotesize \baselineskip 11pt The magnetic field strength $|B|$  with electric field arrows in the plane  at $t=1700 \omega_{\rm pe}^{-1}$ for (upper panels) the $e^{-}-p^{+}$ case and (lower panels) the $e^{\pm}$ case.
Panels (a) and (c) show the electron density in the $x-z$ plane at $y=500\Delta$, and panels (b) and (d) show a cross section in the $y-z$ plane at $x=1200\Delta$.  Color bars (a) 0 - 5.683; (b) 0 - 1.053; (c) 0 - 3.244; (d) 0 - 4.150 \label{fig6}}
\end{figure}

Figure \ref{fig6} shows the magnetic field strength in the $x-z$ plane at $y=500\Delta$ and in a cross section
in the $y-z$ plane at $x=1200\Delta$. 
Figure \ref{fig6}a shows a strong magnetic field in the jet between $500 < x/\Delta < 800$ where the electron-proton jet collimates due to strong toroidal magnetic fields. The magnetic field strength at $x\approx 700\Delta$ indicates a maximum toroidal magnetic field strength at radius $25\Delta<r_{\rm jt}< 100\Delta$.  
The strong toroidal magnetic fields  are also seen in Figs. \ref{fig11}a and \ref{fig12}a.  For the $e^{-}-p^{+}$ case waves generated by the kKHI and the MI are visible in Figure \ref{fig6}b at $x= 1200\Delta$. Strongly growing MI waves are seen at two and five clock (Fig.  \ref{fig6}b).
The layers of concentric rings around the jet seem to be generated by kKHI. For the $e^{\pm}$ case the strongest magnetic fields are located near $x\approx1100\Delta$ where the Weibel instability grows. In the transverse plane at the jet-ambient boundary patterns induced by the MI are visible in Fig. \ref{fig6}d.

\begin{figure}[h!]
\begin{center}
\epsscale{0.9}
\plottwo{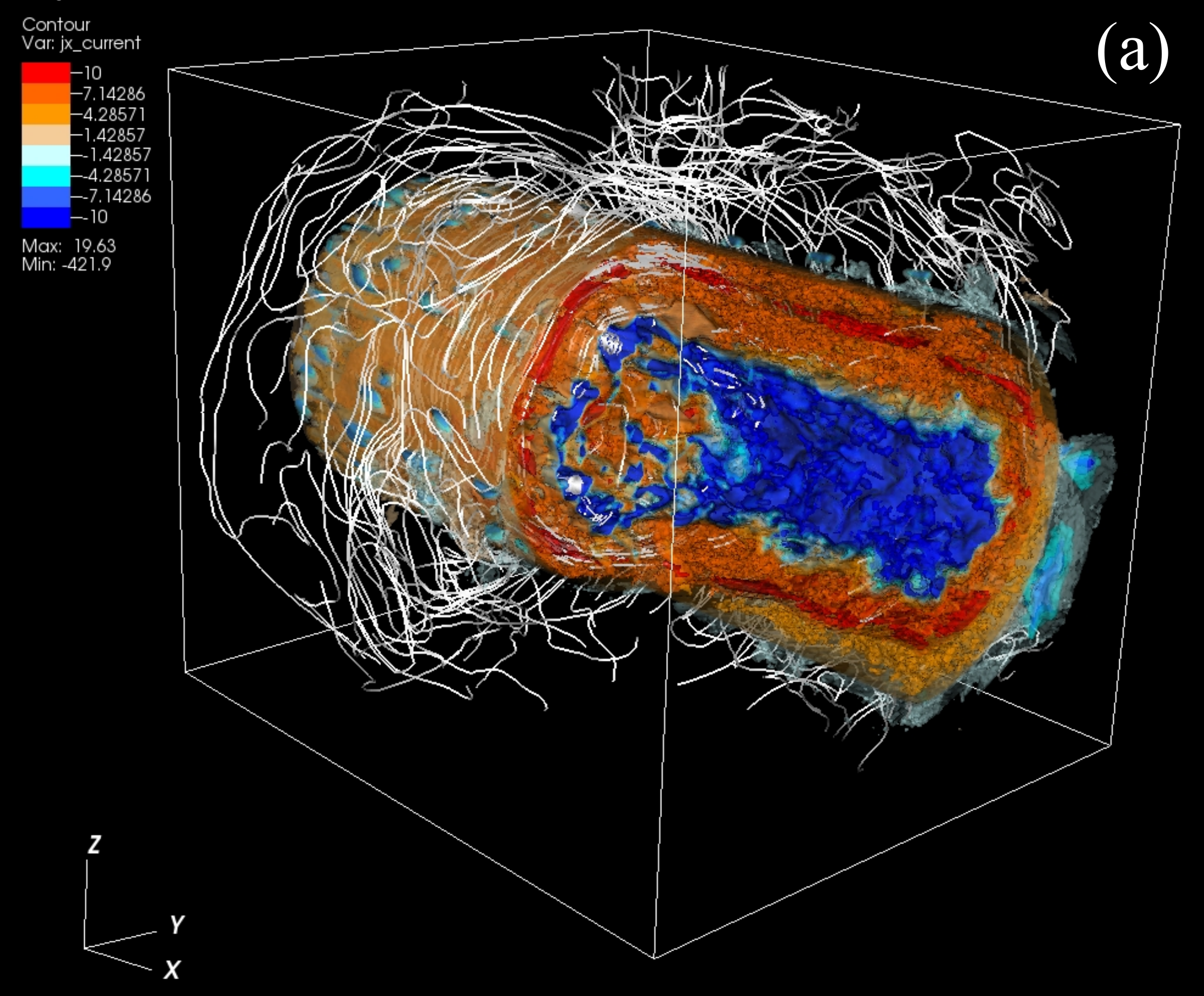}{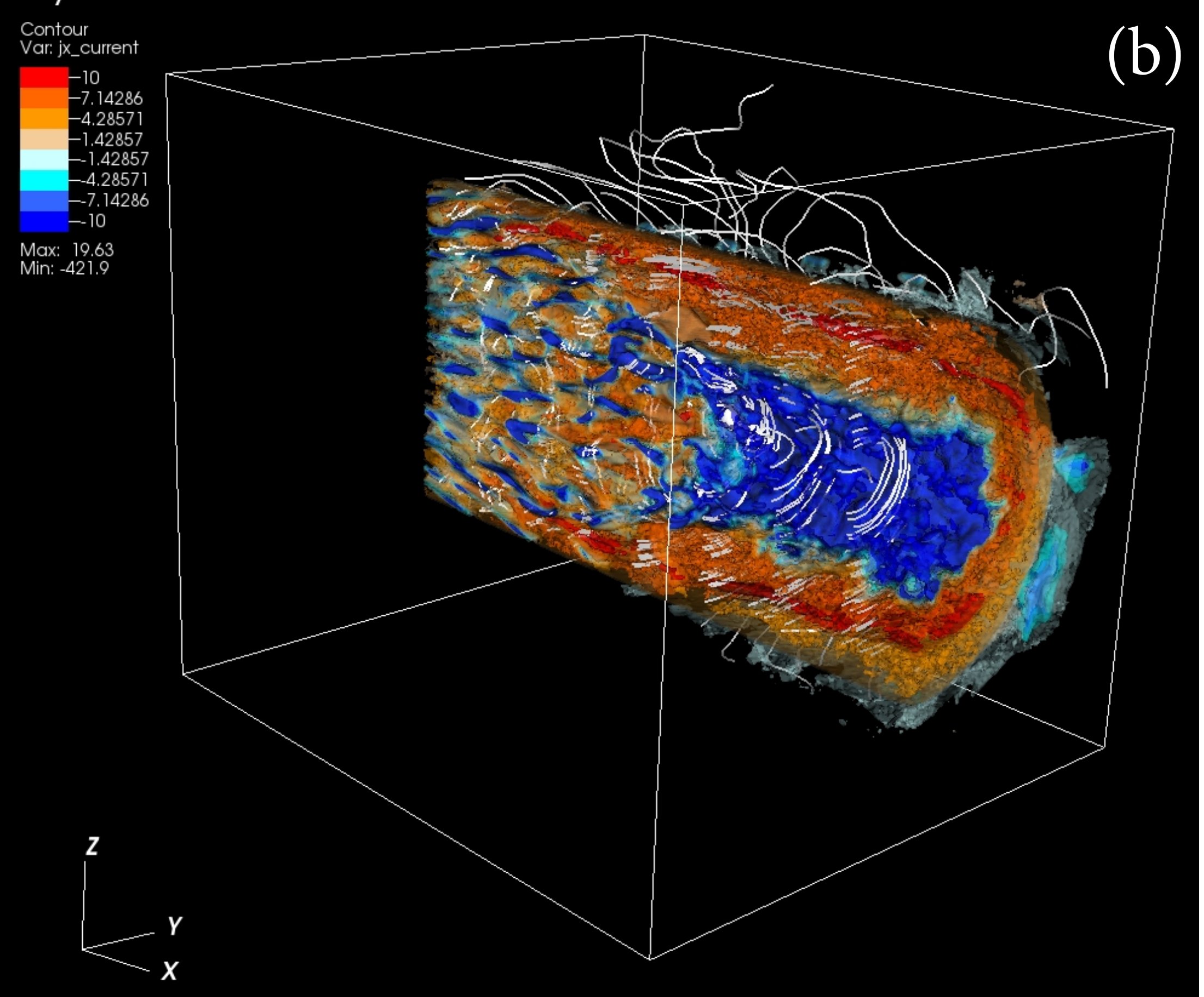}

\vspace*{0.2cm}
\epsscale{0.9}
\plottwo{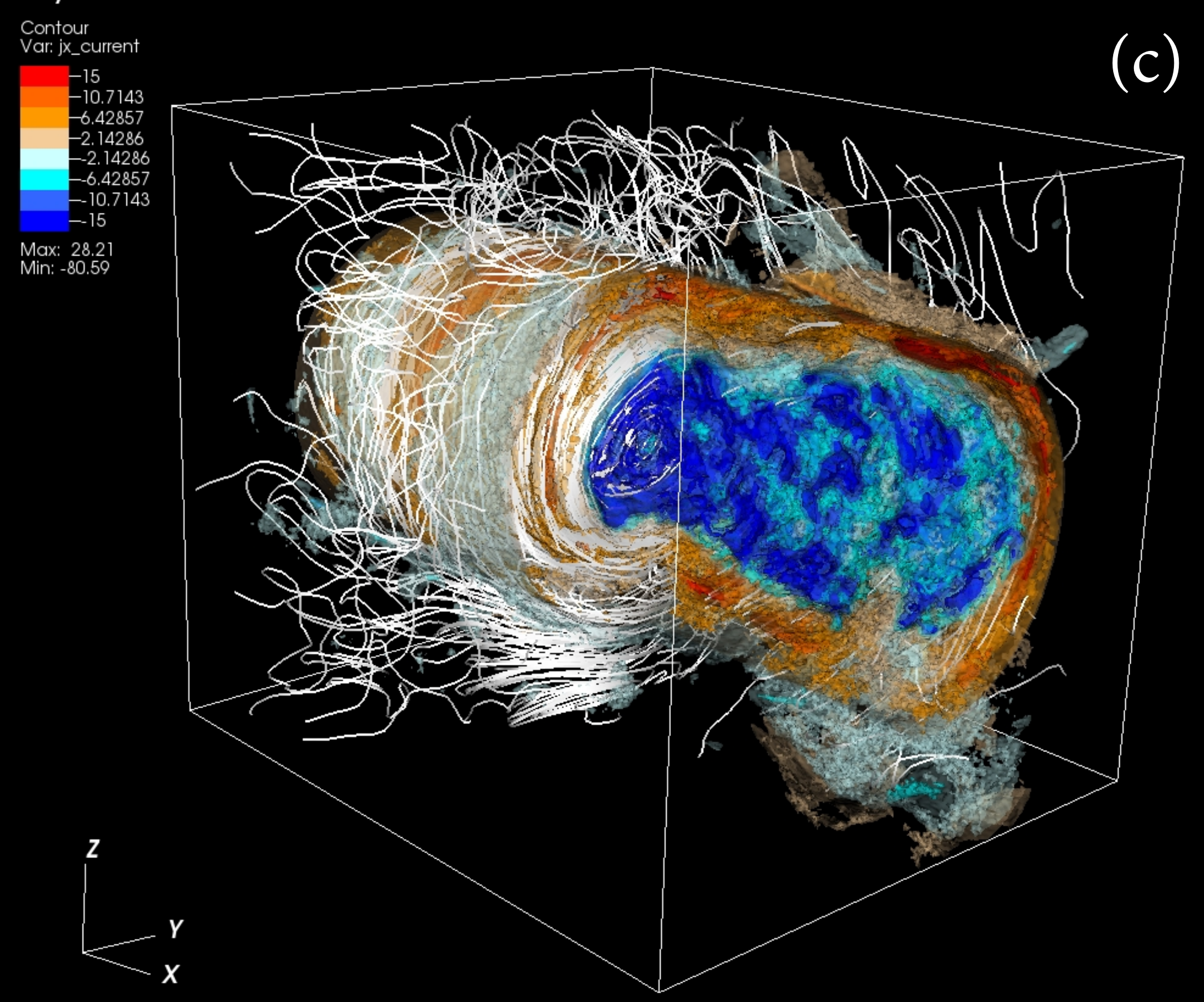}{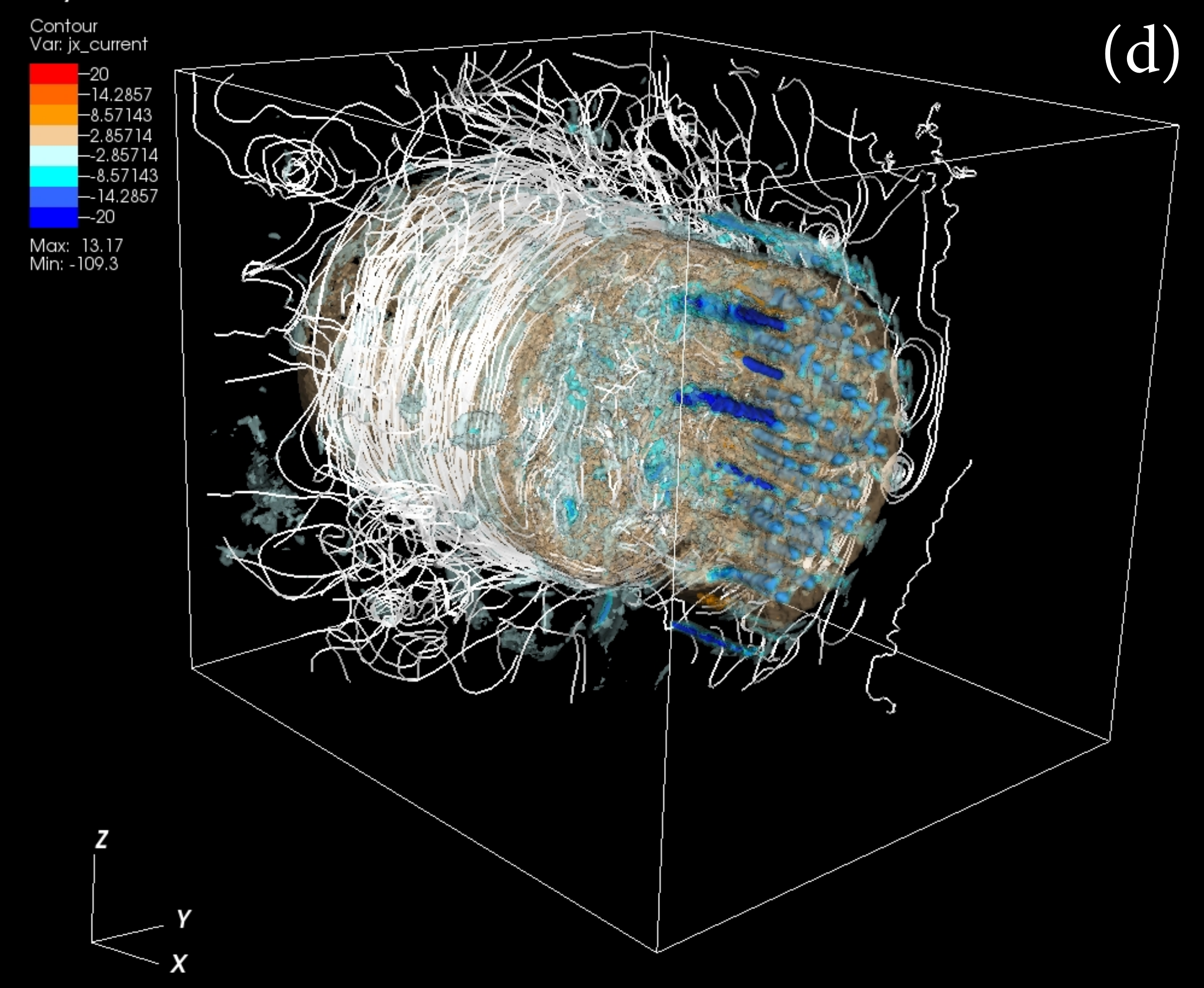}
\end{center}
\vspace*{-0.5cm}
\caption{\footnotesize \baselineskip 11pt 3D structure of the current density component $J_{\rm x}$ clipped at the jet center 
parallel  and perpendicular to the jet for the $e^{-}-p^{+}$ case at $t=1700 \omega_{\rm pe}^{-1}$. White magnetic field lines are shown in the region ($303 < y/\Delta, z/\Delta< 703$).
Panels (a) and (b) show $J_{\rm x}$ in the region $400 < x/\Delta < 800$ where panel (b) shows
an inside region around jet center.   Panels (c) and (d) show $J_{\rm x}$
in the regions $800 < x/\Delta < 1200$, and  $1600 < x/\Delta < 2000$, respectively. Color bars (a) $\pm 10.0$; 
(b)  $\pm 10.0$;  (c)  $\pm 15.0$;  (d) $\pm 20.0$ \label{fig7}}
\end{figure}

Figure \ref{fig7} shows the 3D structure in the $x$ component of the current density in the jet for the $e^{-}-p^{+}$ case near the injection region ($400 < x/\Delta < 800$), in an intermediate region ($800 < x/\Delta < 1200$), and near the jet front ($1600 < x/\Delta < 2000$). 
Collimation occurs near $x\approx 600\Delta$ where proton current filaments merge into the
uniform electron current inside the toroidal proton current shown in Figures \ref{fig7}a and \ref{fig7}b.
This uniform electron current can also be seen in Figure \ref{fig4}a and is associated with the collimated electron density seen in Figure \ref{fig3}a. The collimation is generated by the strong toroidal magnetic field shown in Figure \ref{fig6}a. This strong toroidal magnetic field is generated by the kKHI and/or MI. Based on the
growth rates shown in  Figure 1 in Alves et al. (2015) the MI has much higher growth rate for $\gamma_{\rm jt} = 15$. The structure of the magnetic field lines outside the jet is a signature of the MI. Since MI grows transverse to the jet, and as in the slab model
the magnetic fields become DC when the current filaments merge in the nonlinear stage (see Nishikawa et al. 2014b). 
The strongest toroidal magnetic fields are located inside the jet surface and are hidden by the proton
currents shown in red in Figures \ref{fig7}a and \ref{fig7}b. Figure \ref{fig7}c shows that the collimation relaxes gradually around $x\approx1000\Delta$. Current filaments and nonlinear phenomena at the jet front need further future investigation with a 
fully-evolved shock system.

\begin{figure}[h!]
\begin{center}
\epsscale{0.9}
\plottwo{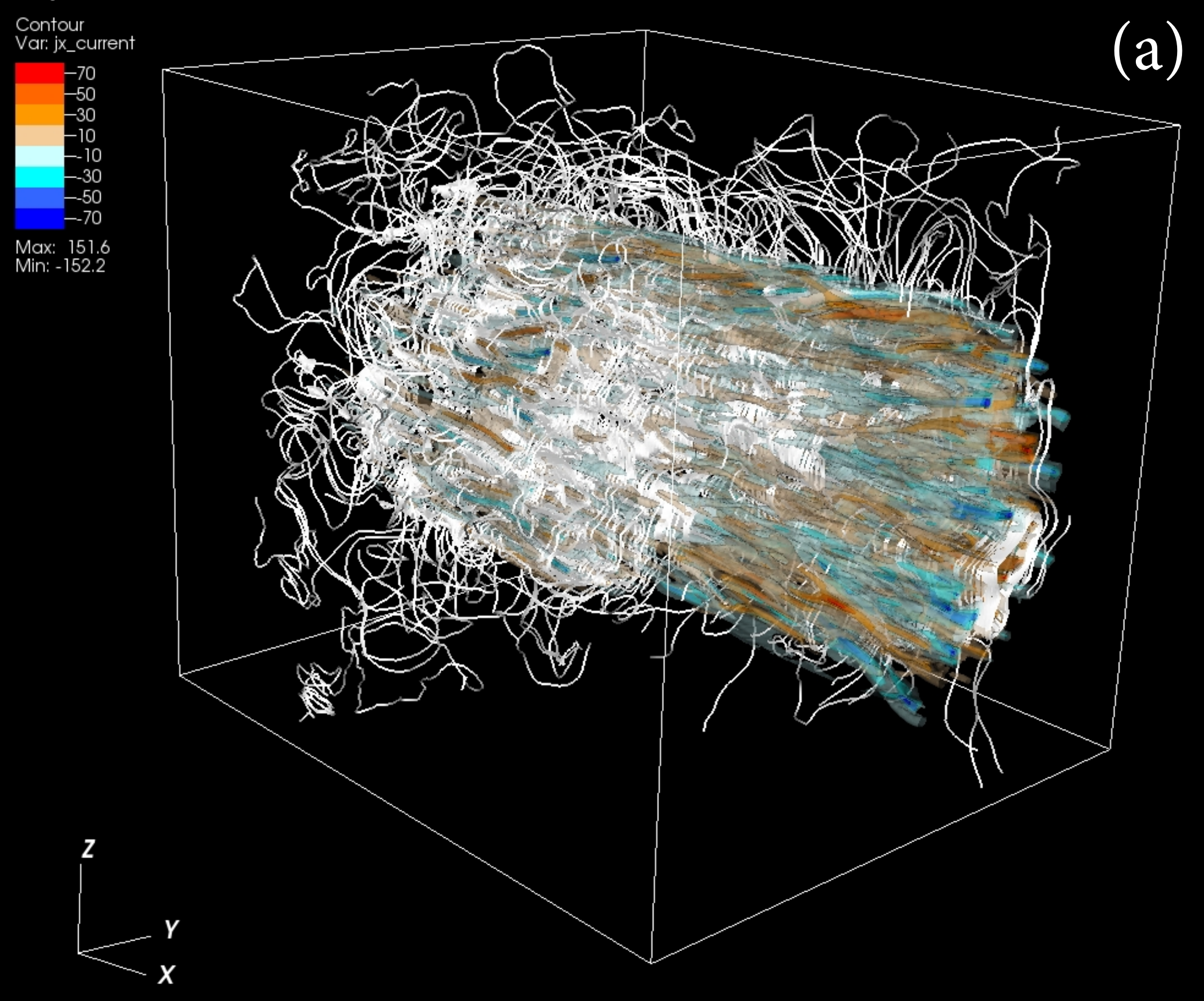}{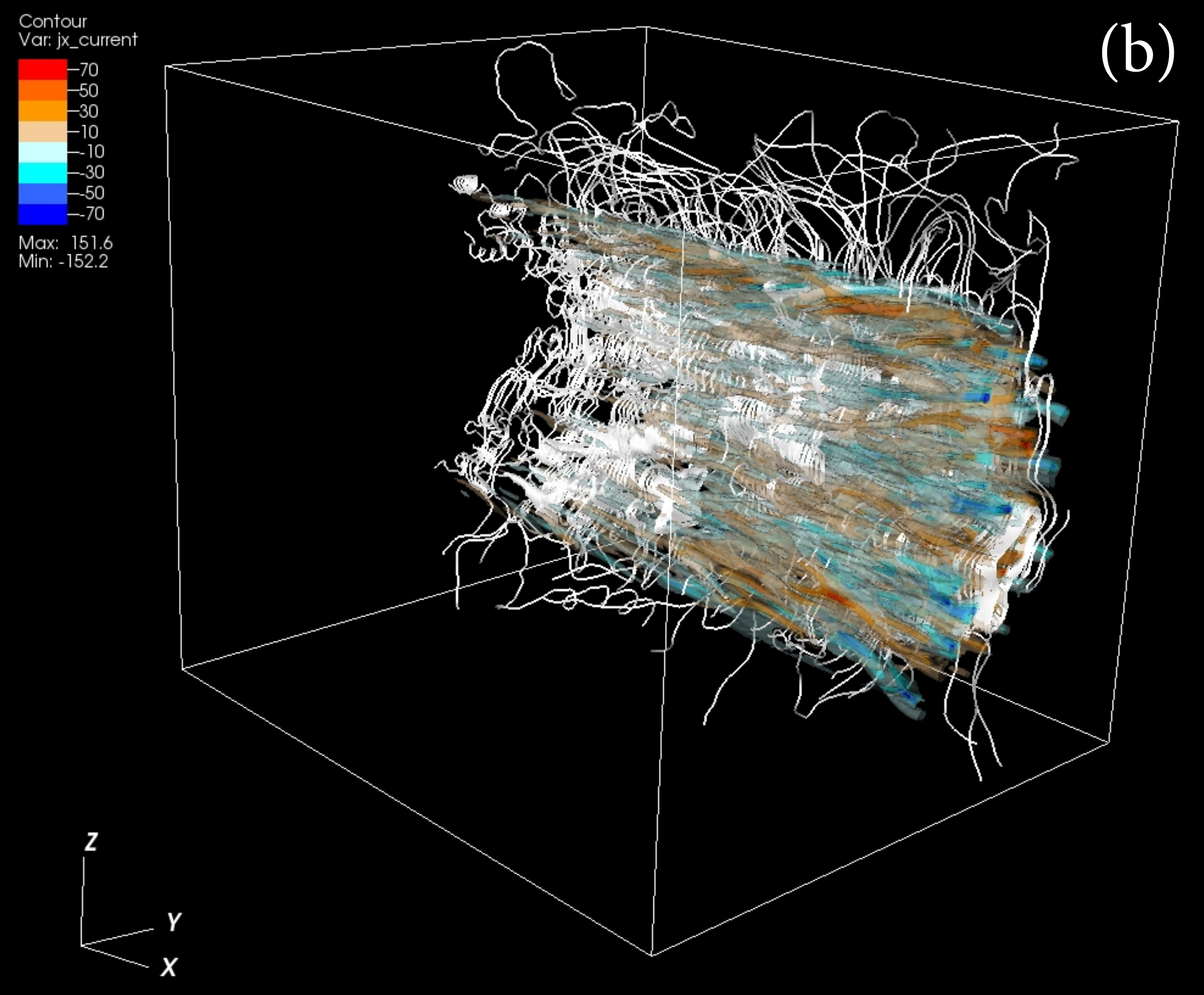}

\vspace*{0.2cm}
\epsscale{0.9}
\plottwo{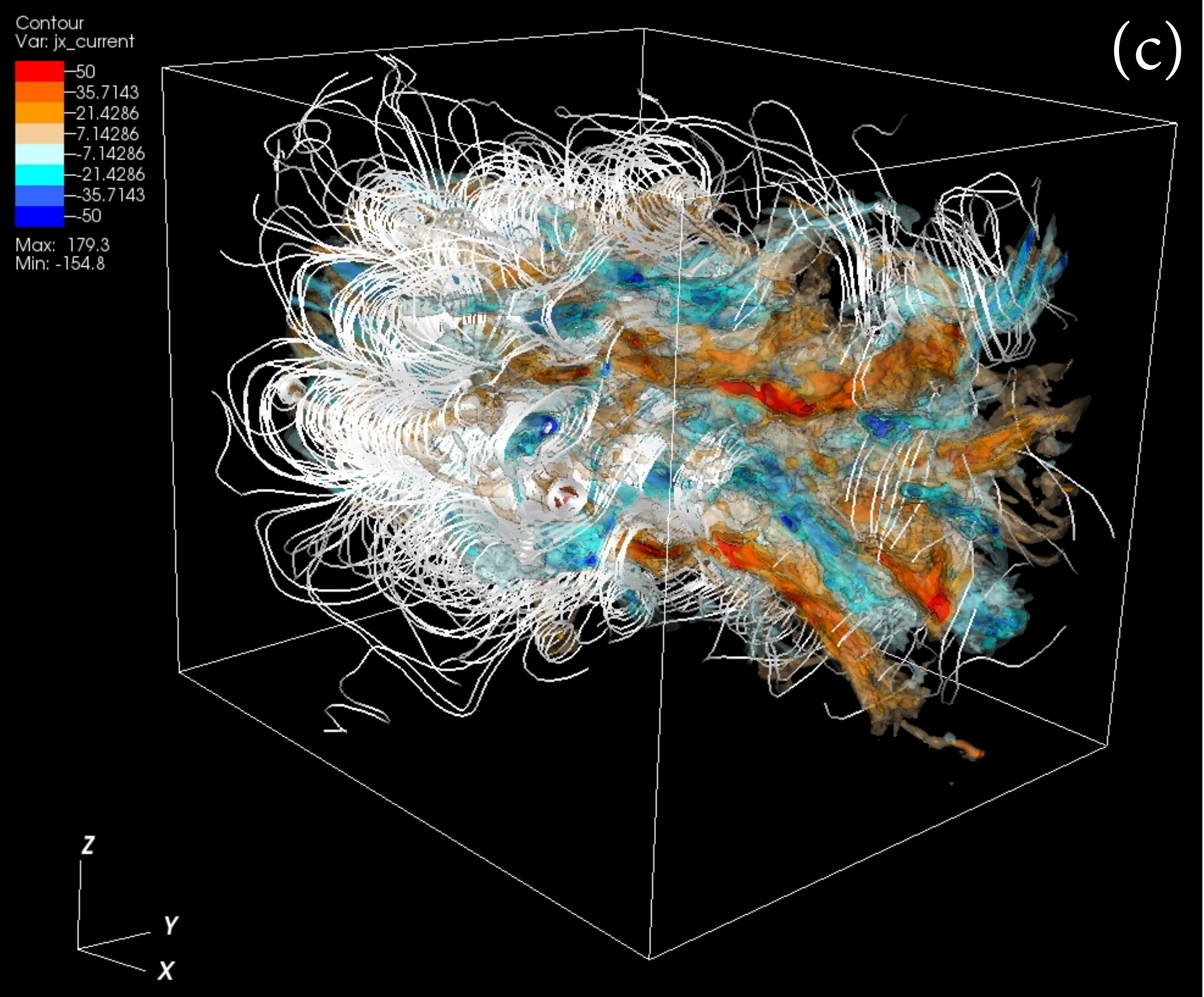}{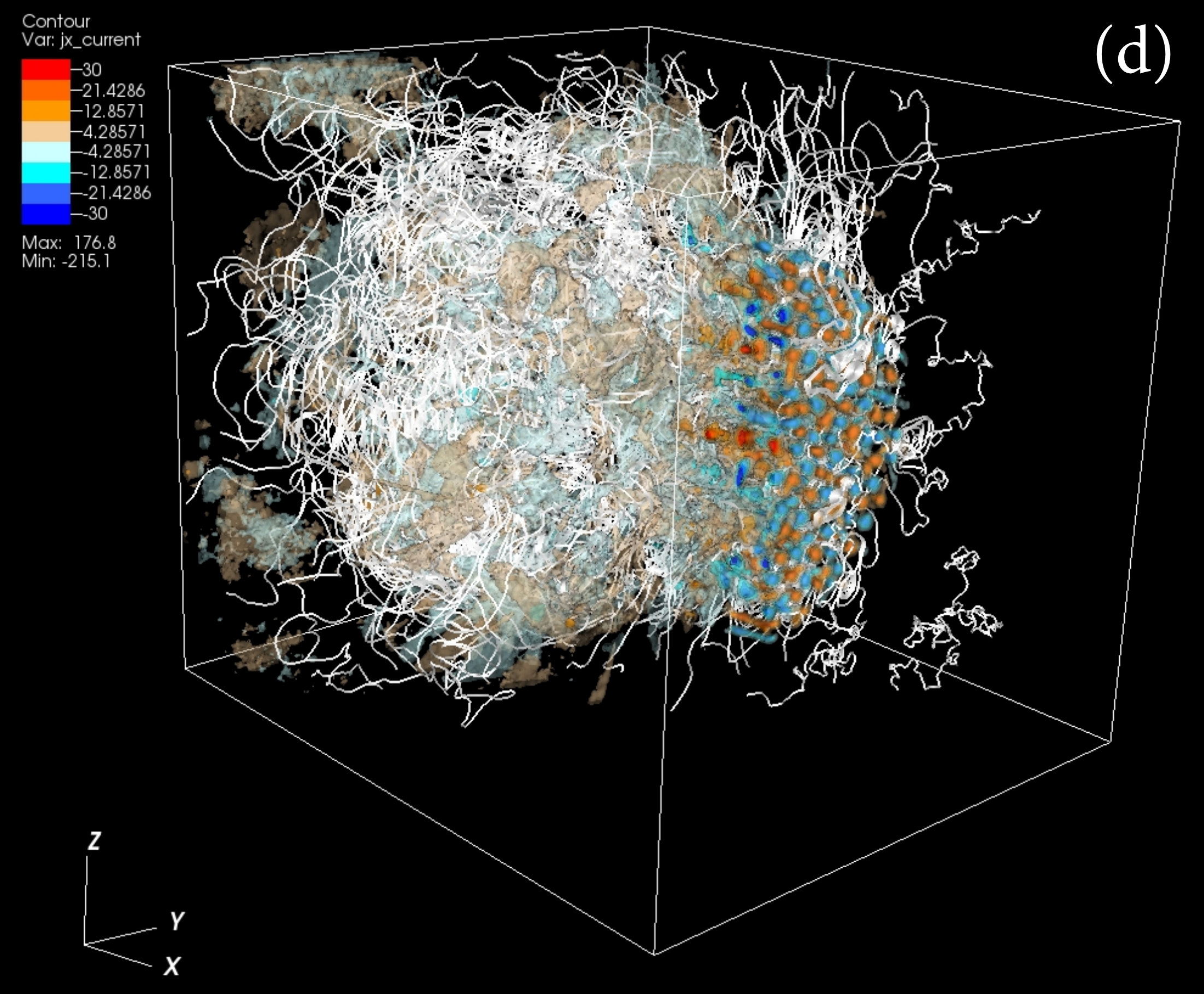}
\end{center}
\vspace*{-0.6cm}
\caption{\footnotesize \baselineskip 11pt 3D structure of the current density component $J_{\rm x}$ clipped at the jet center
parallel  and perpendicular to the jet for the $e^{\pm}$ case at $t=1700 \omega_{\rm pe}^{-1}$. White magnetic field lines are shown in the region ($303 < y/\Delta, z/\Delta< 703$).
Panels (a) and (b) show  $J_{\rm x}$ in the region $400 < x/\Delta < 800$ where panel (b) shows
an inside region around jet center.   Panel s(c) and (d) show $J_{\rm x}$
in the regions $800 < x/\Delta < 1200$, and  $1600 < x/\Delta < 2000$, respectively.   
Color bars (a) $\pm 70.0$; (b)  $\pm 70.0$;  (c)  $\pm 50.0$;  (d) $\pm 30.0$ \label{fig8}}
\end{figure}

Figure \ref{fig8} shows the 3D structure in the $x$ component of the current density in the jet 
for the $e^{\pm}$ case near the injection region ($400 < x/\Delta < 800$), in an intermediate region ($800 < x/\Delta < 1200$), and near the jet front ($1600 < x/\Delta < 2000$). 
\vspace*{-0.cm}
As in a previous electron-positron simulation (Nishikawa et al. 2009a), current filaments grow and in the nonlinear stage are located outside the jet as shown in Figure \ref{fig8}c, and merge and/or dissipate as shown in Figure \ref{fig4}c.  After merging current filaments become weak by $x\approx1300\Delta$. This evolution is very similar to a previous simulation at the same simulation time $t = 1700\omega_{\rm pe}^{-1}$ (e.g., Nishikawa et al. 2009a). 
In the $e^{\pm}$ case, as previously found in the slab model, AC magnetic field patterns are generated at the velocity shear and no collimation occurs.

\begin{figure*}[h!]
\begin{center}
\epsscale{0.9}
\vspace*{-0.2cm}
\plotone{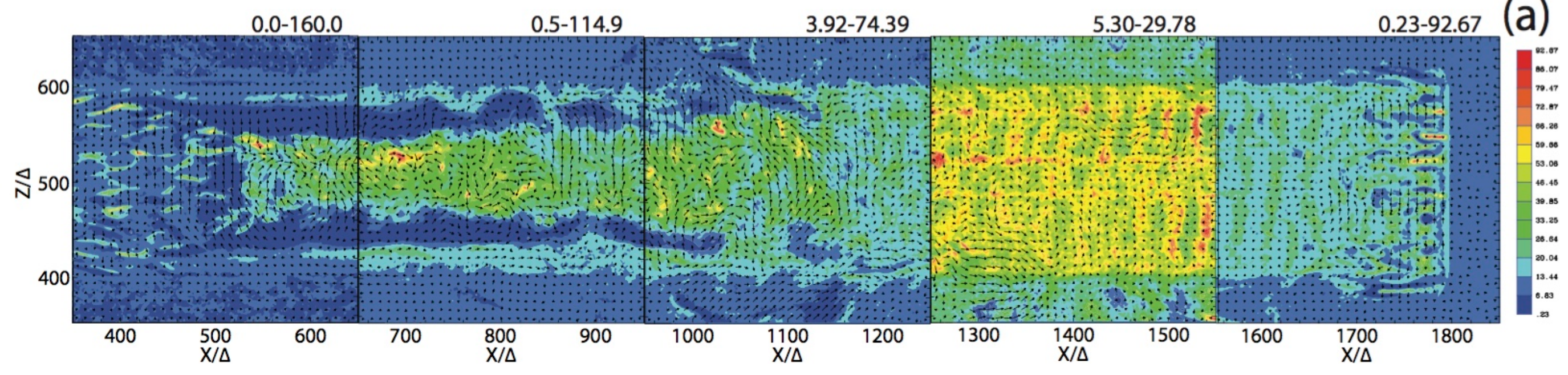} 
\plotone{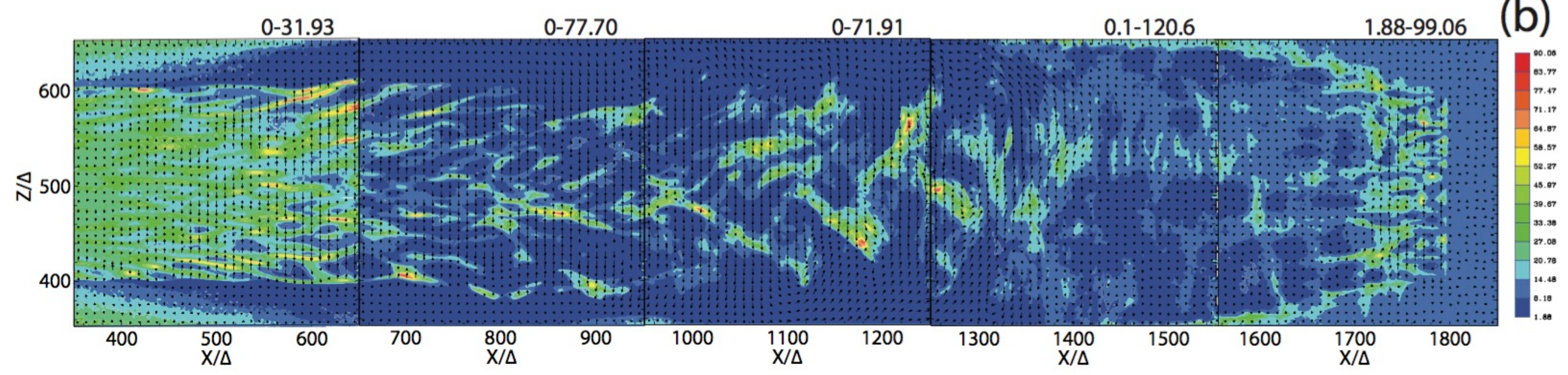} 
\end{center}
\vspace*{-0.6cm}
\caption{\footnotesize \baselineskip 11pt 2D slices of the electron density for (a) the $e^{-}$- $p^{+}$ case and (b) the $e^{\pm}$ case at time $t =  1700\omega_{\rm pe}^{-1}$. Arrows show $B_{\rm x,z}$. The maximum and minimum values of the electron density in each of the 5 jet regions are indicated at upper right.  
 \label{fig9}}
\end{figure*}

Figure \ref{fig9} shows the electron density  in a 2D slice through jet center for the $e^{-}$- $p^{+}$  (upper) and $e^{\pm}$ (lower) at time $t =  1700\omega_{\rm pe}^{-1}$. In order to show the electron density structure in detail, five sections are scaled separately. The minimum and maximum numbers of each panel
are indicated at the right upper corner.  For the  $e^{-}$- $p^{+}$ case the  electron density increases 
at $550 < x/\Delta <900$ due to collimation. The collimation relaxes by $x/\Delta \approx1150$ and the electron density spreads to the original jet width. 
For the $e^{\pm}$ case electron density  filaments are initially generated along the jet and  in the nonlinear stage move outside of the jet as shown in Figures \ref{fig9}b and \ref{fig8}c.
\vspace*{-0.0cm}

\begin{figure*}[h!]
\begin{center}
\epsscale{0.9}
\vspace*{-0.2cm}
\plotone{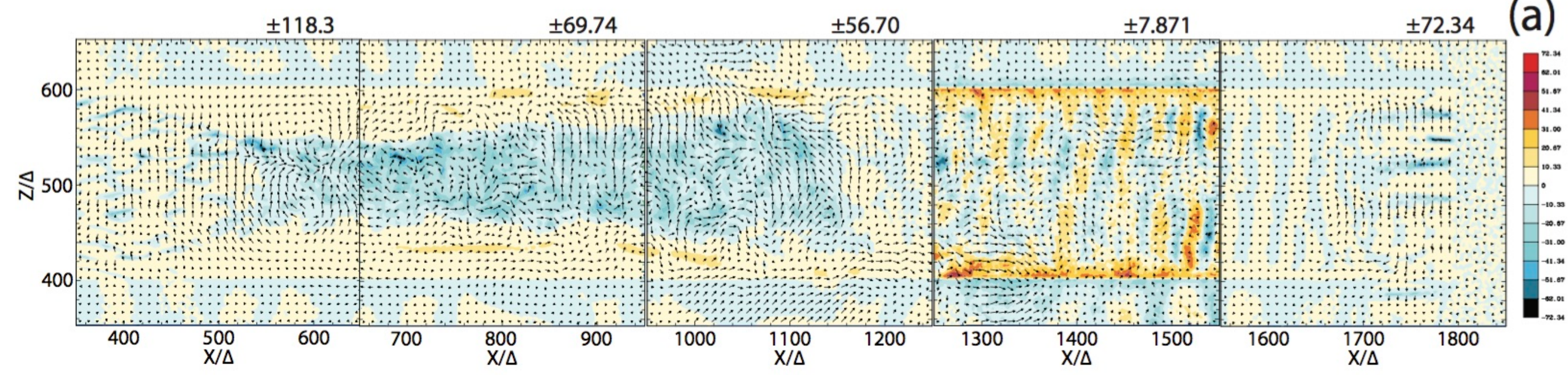} 
\plotone{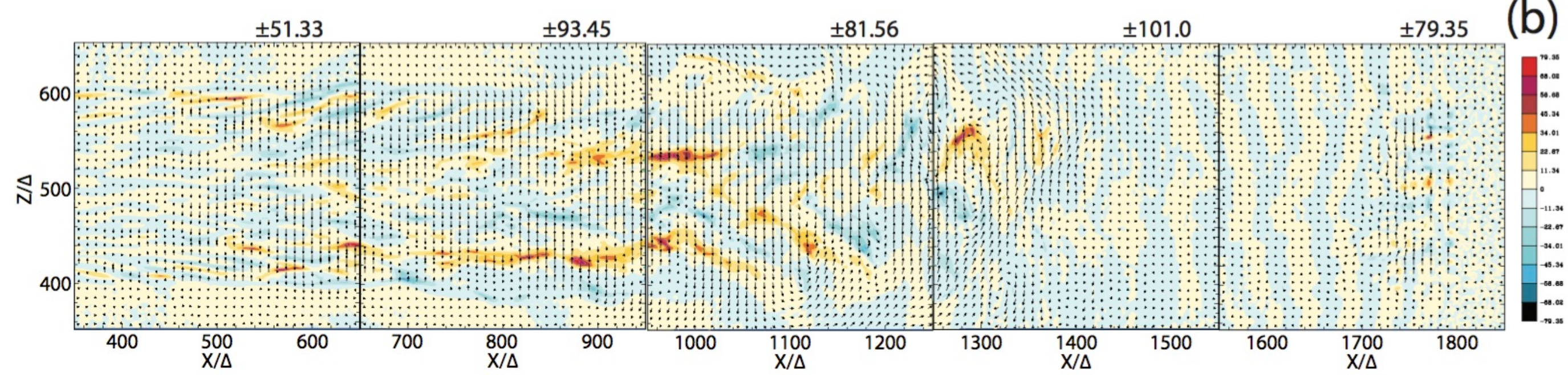} 
\end{center}
\vspace*{-0.6cm}
\caption{\footnotesize \baselineskip 11pt 2D slices of $J_{\rm x}$  for (a) the $e^{-}$- $p^{+}$ case and (b) the $e^{\pm}$ case at time $t =  1700\omega_{\rm pe}^{-1}$. Arrows show $B_{\rm x,z}$. 
The color scale at the right is for only the rightmost panel (region $1550<x/\Delta<1850$). 
The maximum and minimum values in each of the 5 jet regions are indicated at upper right.  
 \label{fig10}}
\end{figure*}

\begin{figure*}[h!]
\begin{center}
\epsscale{0.9}
\vspace*{-0.2cm}
\plotone{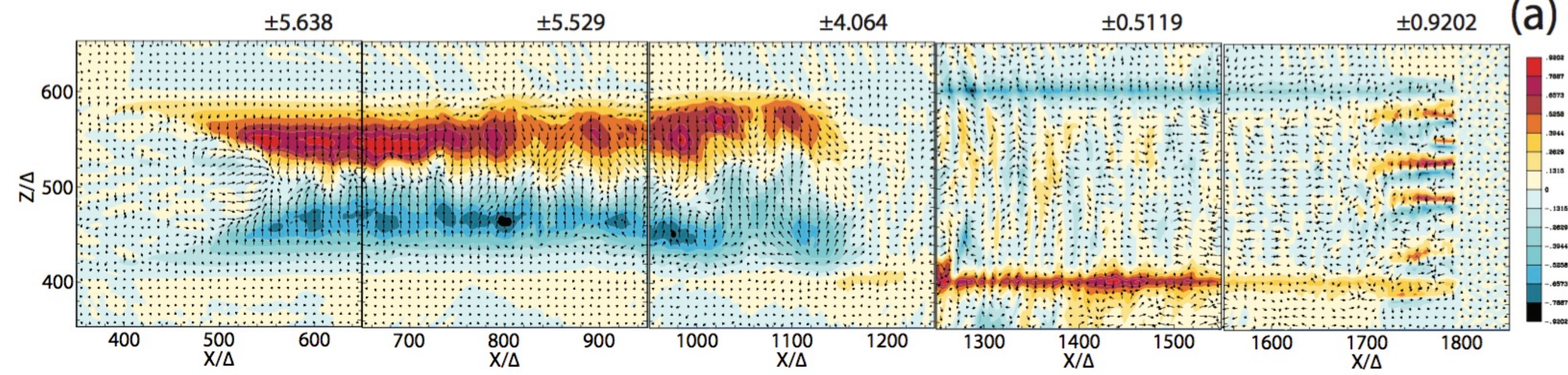} 
\plotone{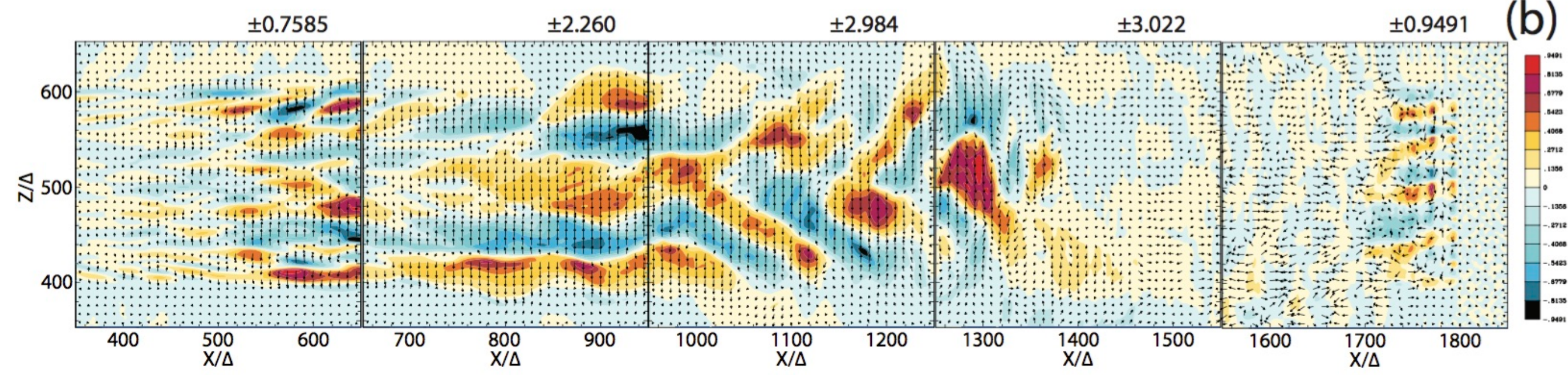} 
\end{center}
\vspace*{-0.6cm}
\caption{\footnotesize \baselineskip 11pt 2D slices of  $B_{\rm y}$  for (a) the $e^{-}$- $p^{+}$ case and (b) the $e^{\pm}$ case at time $t =  1700\omega_{\rm pe}^{-1}$. Arrows show $E_{\rm x,z}$. 
The color scale at the right is for only the rightmost panel (region $1550<x/\Delta<1850$). The maximum and minimum values in each of the 5 jet regions are indicated at upper right.   
 \label{fig11}}
\end{figure*}

\begin{figure*}[h!]
\begin{center}
\epsscale{0.9}
\vspace*{-0.2cm}
\plottwo{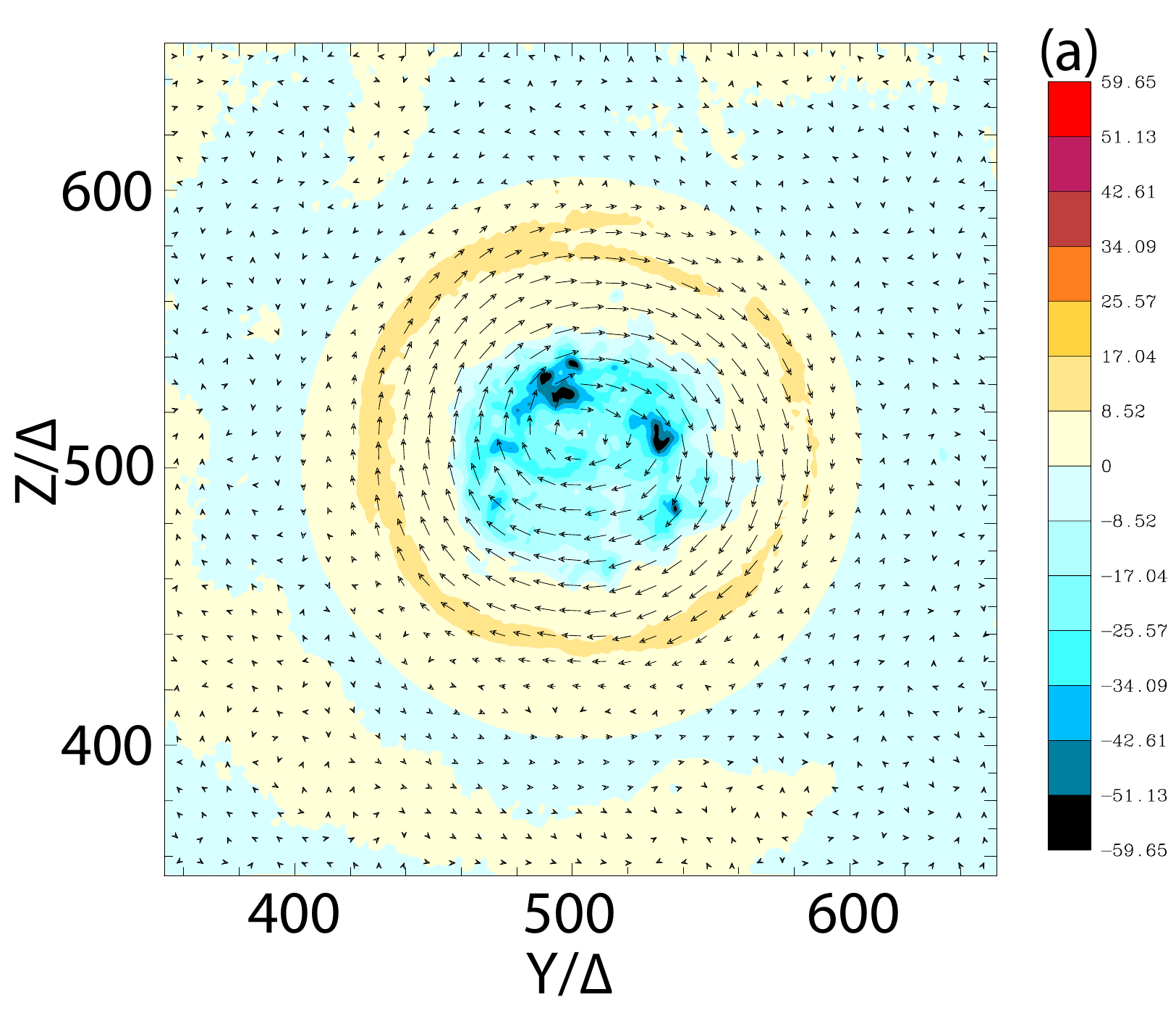}{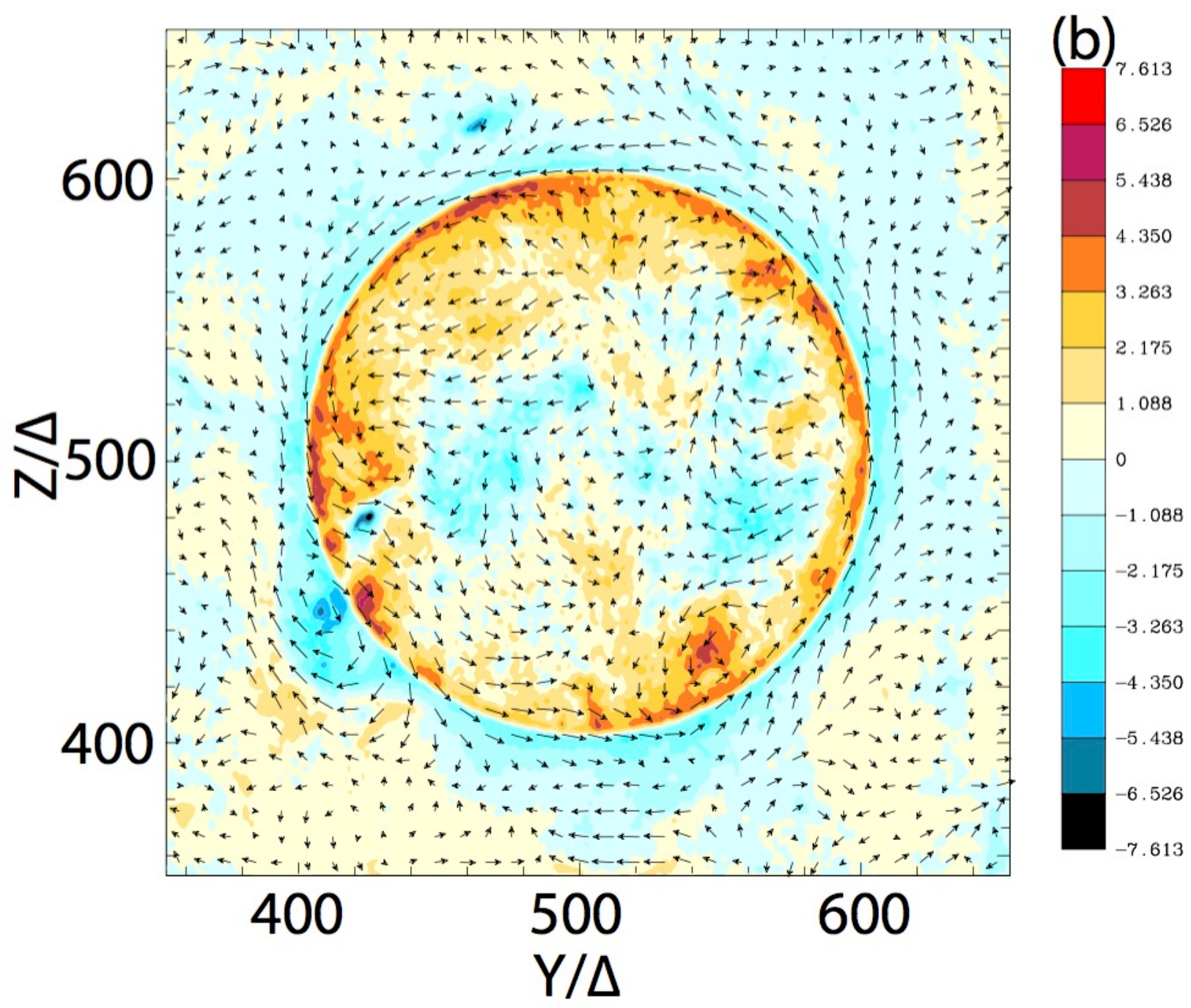} 
\plottwo{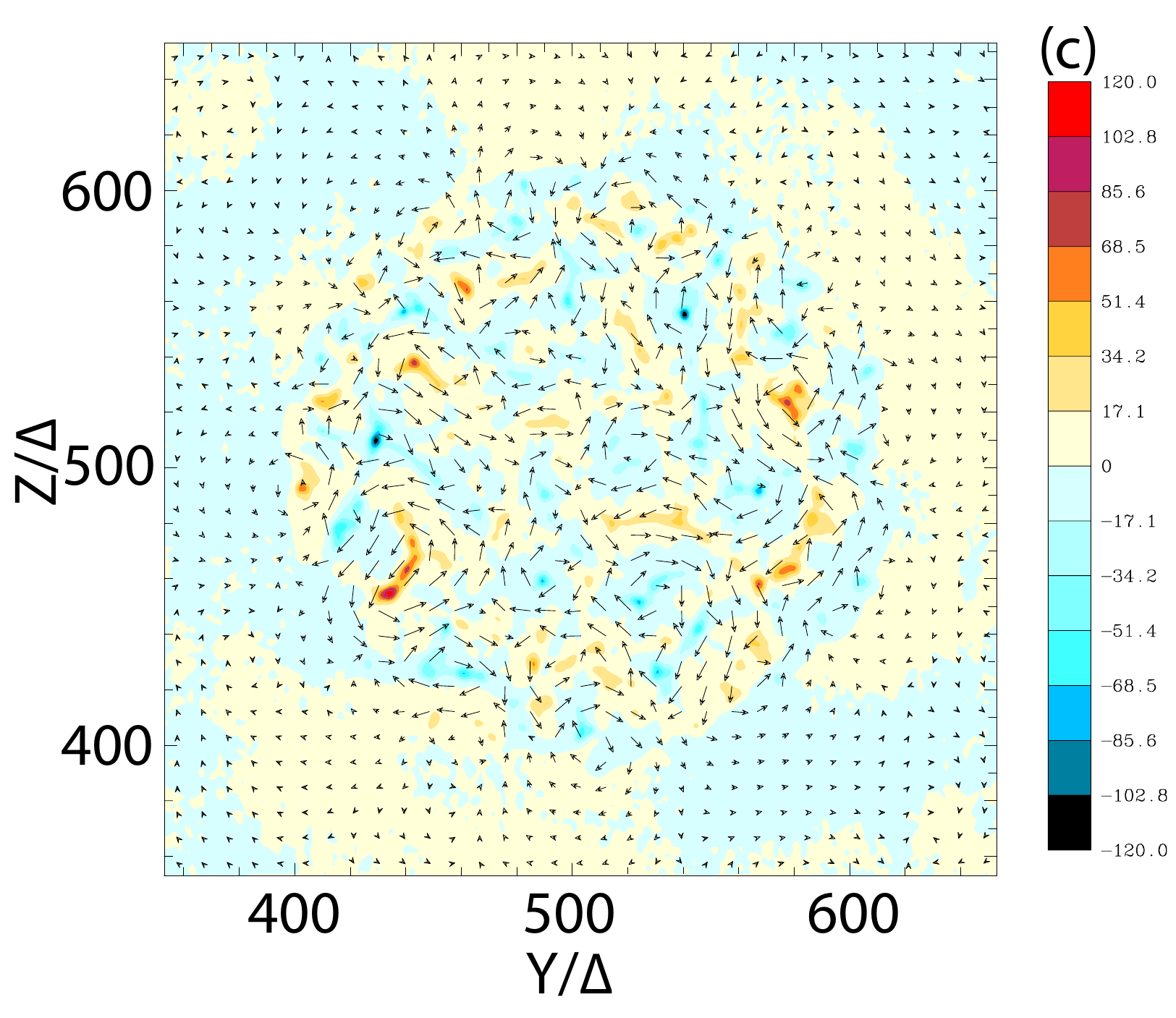}{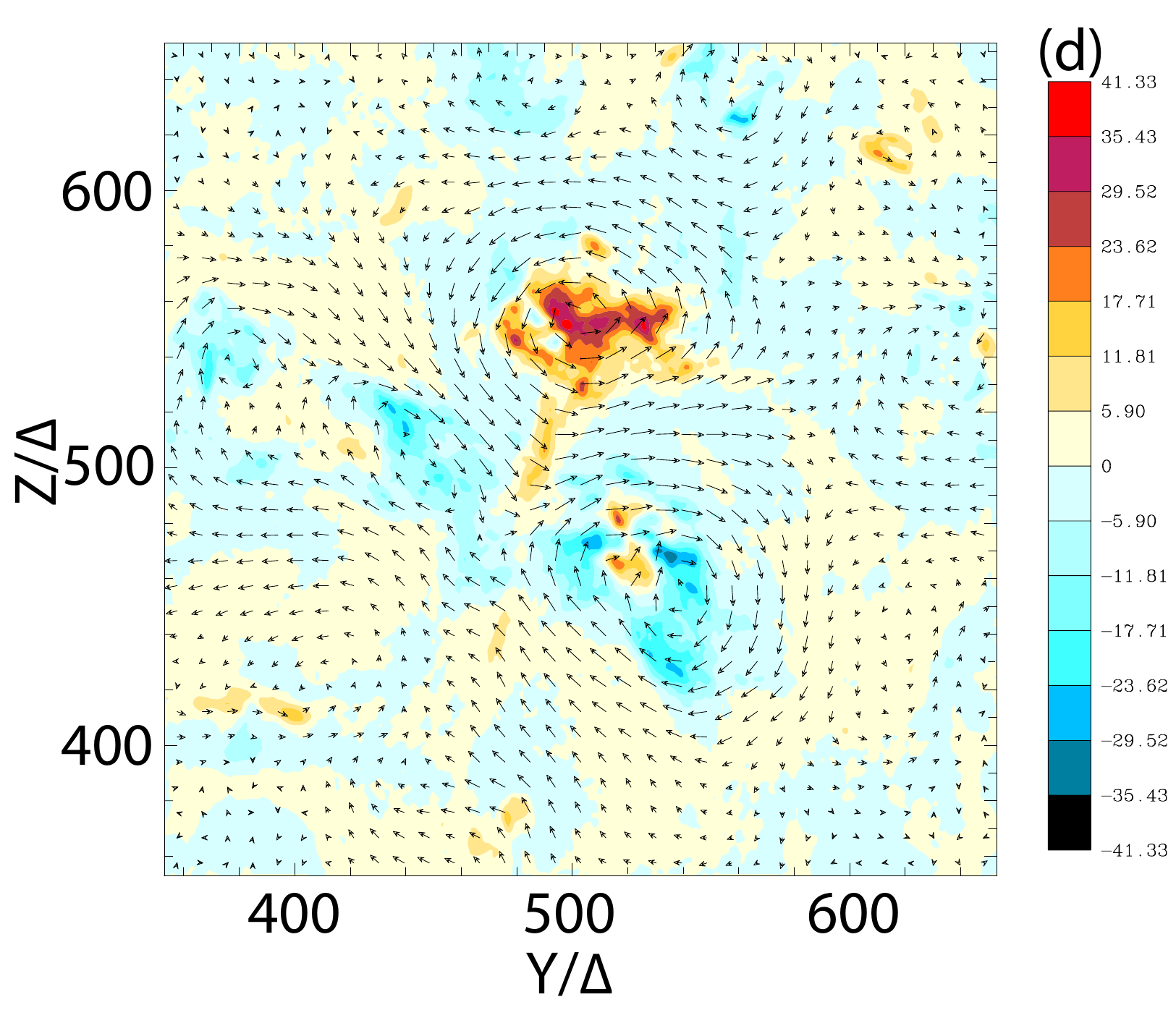}  
\end{center}
\vspace*{-0.6cm}
\caption{\footnotesize \baselineskip 11pt The  $x$ component of  current density $J_{\rm x}$ for (panels a and b) the $e^{-}$- $p^{+}$ case and (panels c and d) the $e^{\pm}$ case at time $t =  1700\omega_{\rm pe}^{-1}$. Arrows show $B_{\rm y,z}$. Panels (a and c) are at $x/\Delta = 700$ and Panels (b and d) are at $x/\Delta = 1300$.  Color bars (a) $\pm 59.65$ ;  (b)  $\pm 7.613$;  (c)  $\pm 120.0$;  (d) $\pm 41.33$
 \label{fig12}}
\end{figure*}

Figure \ref{fig10} shows $J_{\rm x}$ in 2D slices through jet center for the $e^{-}$- $p^{+}$  (upper) and $e^{\pm}$ (lower) cases at time $t =  1700\omega_{\rm pe}^{-1}$. For the  $e^{-}$- $p^{+}$ case  strong electron currents are generated in the center of the jet $550 < x/\Delta <900$ consistent with the collimation shown in Figure \ref{fig9}a. After the collimation relaxes around $x/\Delta \approx1150$ proton currents become dominant at the jet boundary and the electron and proton  currents become layered at $x/\Delta> 1200$.  It should be noted that in the collimated region the magnetic fields shown by the arrows provide rather complex structures. 
For the $e^{\pm}$ case  current filaments are initially generated along the jet and in the nonlinear stage they move out of the jet as shown in Figures {\bf \ref{fig10}b,} \ref{fig11}b and \ref{fig8}c.
\vspace*{-0.0cm}

Figure \ref{fig11} shows 2D slices of $B_{\rm y}$ through the jet center for the $e^{-}$- $p^{+}$  (upper) 
\vspace*{-0.0cm}
and $e^{\pm}$ (lower) cases at time $t =  1700\omega_{\rm pe}^{-1}$. In the  $e^{-}$- $p^{+}$ case strong toroidal magnetic fields collimate the jet. 
This strong toroidal magnetic field is generated by the strong $-J_{\rm x}$  by collimated jet (ambient) electrons 
as shown in Fig. \ref{fig4}a.
After the collimation relaxes around $x/\Delta \approx1150$ the polarity of the toroidal magnetic fields switches from clockwise in the rightmost panels to counter-clockwise in the leftmost panels as viewed from the jet front. The counter-clockwise magnetic fields are generated by the current layer ($+J_{\rm x}$) at the jet boundary as shown in the right two panels in Figure \ref{fig10}a.
Due to the perpendicularly accelerated jet electrons in the collimated region as shown in Fig. \ref{fig5}b, the electrons are expelled from the collimated region and are slowed down as shown in Fig. \ref{fig5}b. Consequently, 
the MI is saturated and the strong toroidal magnetic field disappears and releases the collimation. At the same time, concentrated electron current near the center of the jet brakes and jet electrons expand outward. As shown in Fig. \ref{fig12}b, heavy jet protons maintain the original jet border line and due to the decrease of electrons near the jet boundary $+J_{\rm x}$ current is generated, which generates the counter-clockwise magnetic field. Furthermore 
the patterns of MI are seen near the jet boundary.

For the $e^{\pm}$ case,  filaments of alternating $B_{\rm y}$ are initially generated along the jet  by the current filaments and  at the nonlinear stage they move out of the jet as shown in Figures \ref{fig11}b and \ref{fig8}c.
For both cases at the jet front the strong striped $B_{\rm y}$ components generated by the current filaments
as shown in Figure \ref{fig10}, which have been observed in previous simulations of the Weibel instability 
(Nishikawa et al. 2009a; Ardaneh et al. 2015).

Figure \ref{fig12} shows $J_{\rm x}$ in 2D cross sections at  $x/\Delta = 700$ and $x/\Delta = 1300$
\vspace*{-0.0cm}
for (panels a and b) the $e^{-}$- $p^{+}$ case and (panels c and d) the $e^{\pm}$ case at time $t =  1700\omega_{\rm pe}^{-1}$. The arrows show the magnetic field $B_{\rm y,z}$ in the cross section plane. For the $e^{-}$- $p^{+}$ case  at $x/\Delta = 700$ the toroidal magnetic field is clockwise and generated by the electron current in the center of the jet.  Since electrons are collimated, protons move slightly toward jet center as indicated by stronger $+J_{\rm x}$ located slightly inside the original jet boundary. At $x/\Delta = 1300$ the magnetic field is counter-clockwise and generated by  proton dominated currents slightly inside the  jet boundary. For the $e^{\pm}$ case shown in Figure \ref{fig12}c small current filaments are generated by the kKHI, the MI, and the Weibel instability. The slightly stronger current filaments around the jet boundary are likely  generated by the MI. These small current filaments merge into the two larger current filaments at $x/\Delta = 1300$ shown in Figure \ref{fig12}d.  

\vspace{-0.7cm}
\section{Summary and Discussion}
\vspace{-0.2cm}

We performed  simulations of a relativistic cylindrical jet that is injected into an ambient  plasma in order to investigate a simultaneous velocity shear and shock system, and to study the interaction between shear instabilities (kKHI and MI) and the filamentation Weibel-like instability.  The investigation of the combined processes shows 
very different evolution for $e^{-}$- $p^{+}$  and  $e^{\pm}$ plasma cases. Our jet radius and transverse grid dimensions properly reveal the different evolution associated with the velocity shear instability. However, the grid length is not long enough for the leading edge shock system to fully develop as shown in previous shock generation simulations using a much longer grid (Nishikawa et al. 2009a; Choi et al. 2014; Ardaneh et al. 2015).

The key simulation results for $e^{-}$- $p^{+}$ plasma case are:
\begin{list}{}{\setlength{\leftmargin}{0.5cm} \setlength{\itemindent}{-0.5cm}}
\vspace*{-0.4cm}
\item
$\bullet$ Jet electrons are collimated by strong toroidal magnetic fields generated by the MI.
\vspace*{-0.3cm}
\item
$\bullet$ Electrons are perpendicularly accelerated along with the jet collimation.

\vspace*{-0.3cm}
\item
$\bullet$ Toroidal magnetic field polarity switches from clockwise to counter-clockwise about half way down the jet. 
\end{list}
The key simulation results for the $e^{\pm}$ plasma case are:
\begin{list}{}{\setlength{\leftmargin}{0.5cm} \setlength{\itemindent}{-0.5cm}}
\vspace*{-0.2cm}
\item
$\bullet$ Jet electrons and positrons mix with the ambient plasma.
\vspace*{-0.3cm}
\item
$\bullet$ Magnetic fields around current filaments generated by a combination of kKHI, MI and Weibel instability merge and generate density fluctuations.
\vspace*{-0.3cm}
\item
$\bullet$ A larger jet radius is required to properly simulate the $e^{\pm}$ case since jet and ambient particles mix strongly.
\end{list}

\vspace*{-0.4cm}

Kinetic processes at the velocity shear, i.e., the kKHI and MI, lead to differences between the $e^{-}$- $p^{+}$  and   $e^{\pm}$ cases. Proper spatial behavior can only be  investigated by the simulation setup used in this study. Due to the merging of current filaments generated by the kKHI and the MI, DC magnetic fields are generated outside the $e^{-}$- $p^{+}$ jet. These strong toroidal magnetic fields collimate the jet electrons. 
After the instabilities saturate and relax, the polarity of  the toroidal magnetic field switches from  clockwise to counter-clockwise as shown in Figure \ref{fig11}a. The collimation of jet electrons is accompanied by perpendicular acceleration as shown in Figure \ref{fig5}b. 

For the $e^{-}$- $p^{+}$ case, interaction and magnetic fields do not extend far from the initial
velocity shear surface. For the $e^{\pm}$ cases, interaction and magnetic fields extend farther from the 
initial velocity shear surface although the interaction occurs more inside the jet for
higher jet Lorentz factor. 

Differences in velocity shear magnetic field structure resulting from differences in composition should have
consequences for the appearance of jets in very high-resolution radio imaging. For a simple cylindrical 
geometry velocity shear case, an electron-proton jet would primarily build toroidal magnetic
field at the velocity shear surface. The magnetic field would appear quasi-parallel to the line 
of sight at the limbs of the jet for typical aspect angles $\theta \approx  \gamma^{-1}_{\rm jt}$. In contrast, a
pair-plasma jet would generate sizable radial field components that are only about a factor of two weaker than the toroidal field. The strong electric and magnetic fields in the velocity shear zone will be conducive to particle acceleration as shown in Figure \ref{fig5}b. 

Since the DC field is stronger than the AC field in the electron-proton case, a kinetic treatment is clearly required in order to fully capture the generated field structure (Alves et al. 2012). The generated field structure is important because it may lead to a distinct radiation signature (e.g., Medvedev 2000; Sironi \& Spitkovsky 2009b; Martins et al. 2009; Frederiksen et al. 2010; Medvedev et al. 2011; Nishikawa et al. 2009b, 2010, 2011, 2012).

The evolution of the $e^{\pm}$  jet is very similar to previous jet front shock simulations (e.g., Nishikawa et al. 2009a). Since the jet length is only $1700\Delta$, the leading edge shock system is not yet fully formed, but the current filaments are merged as in the shock simulations. 

The dissipation of a significant fraction of the magnetic energy, e.g., via magnetic reconnection, will naturally result in the appearance of  flares when the accelerated particle beam is directed along the line of sight (Giannios et al. 2009; Komissarov et al. 2009;  Zhang \& Yan  2011; Nalewajko et al. 2011; Cerutti et al. 2012, Granot et al. 2012; Komissarov 2012; McKinney \& Uzdensky 2012; Sironi et al. 2015).
Recently,  Beniamini \& Granot (2015) have studied the expected prompt GRB
emission from magnetic reconnection and compare its expected temporal and spectral
properties to observations. Therefore, it is very important to investigate any signatures of reconnection. For example as shown in Figure \ref{fig10}a in the jet around $x/\Delta = 700$ and in Figure \ref{fig10}b 
around $x/\Delta = 1300$, complicated magnetic field structures may be a signature of reconnection.  Future work will address this possibility.

Current and magnetic structures are very different in the electron-proton and electron positron cases. The differences arise from the different mobilities of protons and positrons. 
The toroidal component of the magnetic field found for the $e^{-}-p^{+}$ case would lead to a stratification in the emission across the jet width due to the dependence of the synchrotron emission with the angle between the magnetic field and the line of sight (e.g., Aloy et al. 2000; Clause-Brown et al. 2011). Furthermore, a toroidal magnetic field would lead to a gradient in Faraday rotation across the jet width, produced by the systematic change in the net line-of-sight magnetic field component (Laing 1981; Blandford 1993). Such gradients in Faraday rotation across the jet width have been searched for through simultaneous multifrequency polarimetric VLBI observations of AGN jets. The first detection of a Faraday rotation gradient was reported by Asada et al. (2002) based on VLBA observations of 3C273. Following this initial detection other authors have claimed similar gradients in Faraday rotation in other AGN jets (e.g., Gabuzda et al. 2004; G\'omez et al. 2008; O'Sullivan \& Gabuzda 2009). Hovatta et al. (2012) have studied a sample of 191 extragalactic radio jets observed within the MOJAVE program, finding significant Faraday rotation gradients in CTA102, 4C39.25, 3C454.3, and 3C273; among these sources, the most clear evidence was again obtained for 3C273, for which these authors have also found variations in the Faraday rotation screen over a timescale of three months, suggesting some internal (within the jet emitting region) Faraday rotation. Furthermore, a sign change from positive to negative Faraday rotation was observed in 3C273, that resembles the change in polarity of the toroidal field found in our electron-proton case.

More recently, G\'omez et al. (2015) have investigated the Faraday rotation screen in BL Lacertae through space-VLBI observations with the RadioAstron mission. These observations, achieving an angular resolution of $\sim$20 uas (the highest obtained to date), reveal a gradient in Faraday rotation and polarization vectors as a function of position angle with respect to the core, suggesting the presence of a large-scale helical magnetic field threading the jet in BL Lacertae.

Additional work is required to explore the observational signatures of the different magnetic field properties expected for the $e^{-}-p^{+}$  and $e^{\pm}$ cases. The resulting magnetic field structures are different enough to yield distinctive polarizations in VLBI observations of AGN jets. 
Toroidal magnetic fields inside and outside the electron-proton jet may contribute to circular polarization (CP).
VLBI observations in the MOJAVE sample have revealed CP in about 15\% of the observed sources, with a level of CP of the order of 0.5\% of the local total intensity (Homan \& Lister 2006). Subsequent observations by Gabuzda et al. (2008) and Vitrishchak et al. (2008) of the same sources for which CP was observed in the MOJAVE sample show no disagreement in the sign of the CP, which supports the idea that the CP is indeed related to the intrinsic composition and magnetic field structure of AGN jets. Although a pair plasma cannot generate intrinsic CP, the CP obtained by Wardle et al. (1998) through VLBI observations of 3C279 is explained if the CP is produced by Faraday conversion.  Faraday conversion requires a low energy cutoff in the electron energy distribution and a pair plasma is required to avoid the jet carrying more kinetic energy than observed.

In these simulations, helical magnetic fields are not included. This also is a major difference with RMHD simulations which have a helical magnetic field. Therefore the results in this study may not cover all aspects of real extragalactic jets. Nevertheless, the phenomena in this PIC simulation study are based on the kinetic effects such as kKHI and MI which are not included in RMHD simulations. For example, hot-spots in FRII jets could be explained due to particle (electron) acceleration due to kinetic instabilities. However, the size of jets in this study is very small compared to the real jets, therefore the macroscopic dynamics of jets may not be included fully. Still, it is fruitful to compare these results with parsec scale jets, but careful considerations need to be taken. We plan to perform synergistic simulation studies with RMHD, which will be reported in a separate report.

In our future simulations we will inject jets with a helical magnetic field like that implemented in Markidis et al.\ (2013).   This configuration will allow investigation of the effect of helical magnetic fields on growth of the Weibel instability, the kKHI, and MI as well as to whether an MHD-like kink and/or a KHI velocity shear driven helical twist occurs.

\acknowledgments

This work is supported by NSF AST-0908010, AST-0908040, 
NASA-NNG05GK73G, NNX07AJ88G, NNX08AG83G, 
NNX08AL39G, NNX09AD16G, NNX12AH06G, NNX13AP-21G, 
and NNX13AP14G grants. The work of J.N. has been
supported by Narodowe Centrum Nauki through research
project DEC-2013/10/E/ST9/00662. 
Y.M. is supported by the ERC Synergy Grant ``BlackHoleCam - Imaging 
the Event Horizon of Black Holes''  (Grant No. 610058).  M.P. acknowledges 
support through grant PO 1508/1-2 of the Deutsche Forschungsgemeinschaft. 
Simulations were performed using Columbia and Pleiades facilities at NASA 
Advanced Supercomputing (NAS), and using Kraken and Nautilus at The National 
Institute for Computational Sciences (NICS), and Stampede at The Texas Advanced 
Computing Center, which are supported by the NSF. This research was started during the program 
``Chirps, Mergers and Explosions: The Final Moments of Coalescing Compact Binaries'' 
at the Kavli Institute for Theoretical Physics, which is supported by the National Science 
Foundation under grant No. PHY05-51164. The first velocity shear results using 
an electron-positron plasma were obtained during the Summer Aspen workshop
``Astrophysical Mechanisms of Particle Acceleration and Escape from the Accelerators'' 
held at the Aspen Center for Physics (2013 September 1-15).



\end{document}